\documentclass[prd,nofootinbib,eqsecnum,final]{revtex4}
\pdfoutput=1 
\usepackage{graphicx,color}
  \usepackage{bm}
   \usepackage{amsmath}
    \usepackage{amssymb}
     \usepackage{pifont}
      \usepackage{simplewick}
\usepackage{tikz}
\usepackage[most]{tcolorbox}
\usepackage{rotating}
\usepackage[normalem]{ulem}

\newcommand{\nn}{\nonumber}

\newcommand{\const}{\text{const.}}

\newcommand{\ot}{\leftarrow}

\renewcommand{\(}{\left(}
\renewcommand{\)}{\right)}
\renewcommand{\[}{\left[}
\renewcommand{\]}{\right]}

\renewcommand{\vec}[1]{\bm{#1}}

\begin{document}
\title{
Systematic analysis of double-scale evolution
}

\author{Ignazio Scimemi}
\affiliation{Departamento de F\' isica Te\'orica, \\ Universidad Complutense de Madrid,
Ciudad Universitaria, \\ 28040 Madrid, Spain}
\author{Alexey Vladimirov}
\affiliation{Institut f\"ur Theoretische Physik, \\ Universit\"at Regensburg,\\
D-93040 Regensburg, Germany}

\begin{abstract}
Often the factorization of differential cross sections results in the definition of fundamental hadronic functions/distributions which have a double-scale evolution,  as provided  by a pair of coupled equations. 
Typically, the two scales are the renormalization and rapidity scales. 
The 
two-dimensional structure of their evolution  is
the object of the present study.
 In order to be more specific, we consider the case of the transverse momentum dependent  distributions (TMD). Nonetheless, most of our findings can be used with other double-scale parton distributions. On the basis of  the two-dimensional structure of TMD evolution, we formulate the general statement of the $\zeta$-prescription introduced in \cite{Scimemi:2017etj},
  and we define an optimal TMD distribution, which is a scaleless model-independent universal non-perturbative function. Within this formulation the non-perturbative definition of the distribution is disentangled from the evolution, which clarifies the separation of perturbative and non-perturbative effects in the phenomenology. A significant part of  this work is devoted to the  study of the effects of truncation of perturbation theory
on the double-scale evolution. We show that within truncated  perturbation theory the solution of evolution equations is ambiguous and  this fact generates extra uncertainties within the resummed cross-section. The alternatives to bypass this issue are discussed. Finally, we discuss the sources and distribution of the scale variation uncertainties.

\end{abstract}
\maketitle

\section{Introduction}

The factorization of differential cross sections allows to isolate well-defined hadronic matrix elements which include the information coming from the low-energy parton interactions. The modern factorization theorems define and operate with  multi-variable parton distributions, such as Transverse Momentum Dependent  Distributions (TMD), jet-functions, double parton distributions, etc. Typically these distributions are an outcome of  double-factorization procedures and depend on two factorization scales. The dependence on these scales is dictated by evolution equations, that are often coupled. Therefore, we \textit{de facto} deal with a two-dimensional differential system of evolution equations. Despite  that this fact is well known, it does not seem to have received the sufficient attention in practical phenomenological applications. The double-scale evolution and its consequences on phenomenology are the main object of this work.

In order to make the discussion more specific we concentrate on the TMD distributions and their evolution. Nevertheless and before entering into the details, we would like to remark that the majority of results of the present work is general, and appears every time one considers some distribution with a double-scale evolution. In this sense the discussion of our work can also be valid, with due re-arrangements, in a more general context. The TMD factorization case, discussed here, is {\it per se} important because it is part of the description of important processes like Drell-Yan (DY), vector/scalar boson production and semi-inclusive deep inelastic scattering (SIDIS).


The theoretical definition of TMDs and their properties has been provided in several studies and the list of the most recent  works includes \cite{Collins:2011zzd,GarciaEchevarria:2011rb,Echevarria:2012js,Echevarria:2014rua,Gaunt:2014ska,Becher:2010tm,Chiu:2012ir,Mantry:2010bi}. Currently,
there are several published computer codes that are based on TMD factorization and include higher order perturbative QCD information for low energy Drell-Yan and vector boson production~\cite{Scimemi:2017etj} or that are more specific for vector boson production~\cite{Catani:2009sm,Bozzi:2010xn,Catani:2015vma,Becher:2011xn}. The TMD distributions appear both as initial and final state hadronic matrix elements and are universal, in the sense that they can be extracted in different hadronic processes and in fact they are a central part of the EIC program~\cite{Accardi:2012qut}. Apart from the theoretical definition and consistency of TMD distributions, their actual implementations present a series of problems which is receiving exceptional attention now because of the amount and precision of the present and forthcoming data. Here we propose an optimal realization of the TMD distributions with which it is expected to obtain a better control of theory uncertainties and a simple practical implementation. 

The double-scale evolution of TMD is created by different regimes of field dynamics. One scale is the standard renormalization scale of ultraviolet (UV) logarithms and  another is the rapidity renormalization scale connected to the related divergences. Such a structure was already observed long ago~\cite{Collins:1981uk} and its relevance has been remarked also in the recent formulations of factorization theorems and TMD definitions (see e.g.~\cite{Collins:2011zzd,GarciaEchevarria:2011rb,Chiu:2012ir,Echevarria:2016scs}). The two scales should be treated independently and are equally important for the final computation of cross sections. A similar structure is observed in many modern applications where soft gluon interactions are factorized, e.g. event shapes ~\cite{Berger:2003pk,Stewart:2013faa}, $p_T$-resummation ~\cite{Stewart:2013faa,Ebert:2016gcn}, multi-parton scattering ~\cite{Diehl:2011yj,Vladimirov:2016qkd,Vladimirov:2017ksc}.

There are two important topics for phenomenology  that are directly related to TMD evolution. These are the minimization of theory uncertainties in the evolution and the selection of the best scales for the distribution definitions. Both these topics are problematic, and should be positively resolved by a critic analysis of double-scale nature of TMD evolution.
There is an additional issue that possibly does not damage the prediction power of the approach, but it seriously affects our understanding of the physical picture of hadrons. 
The problem consists in  the correct disentangling of perturbative and non-perturbative effects in the TMD factorization formula. In fact, the traditional choice of scales mixes up the parameters of TMD distributions and the scales of evolution, rendering unclear the interpretation of the distribution and bringing undesired dependence on the perturbative order into the model of TMD distribution.

The theoretical uncertainty of the factorized cross-section is produced by the truncation of the strong coupling perturbative series. Despite the fact that the TMD evolution is known up to the third order in the strong coupling expansion~\cite{Li:2016ctv,Vladimirov:2016dll}, the error coming from the evolution is the largest among all other theoretical inputs, as it has been shown in ref.~\cite{Scimemi:2017etj}. In this work we demonstrate that the theoretical uncertainty of evolution is originated by the  combination of two effects: the actual uncertainty in the next higher order perturbative correction, and the ambiguity of evolution procedure. Therefore, a part of the error-band is fictitious, in the sense that it is produced by a poor comprehension of the double-scale evolution, rather then the lack of perturbative information. Some comments of this effect can be found in the literature (see e.g.~\cite{Collins:2011zzd,Chiu:2012ir}) although they have not been the central topic of any study. In our attempt to cover this gap here, we motivate this statement and we show explicitly that the numerical effect of the evolution ambiguity is huge and, counterintuitively, the error caused by the truncation is larger for larger energies. The ambiguity is not entirely cured by the increase of the perturbative order, and can have even more dramatic consequences on TMD phenomenology. We cite here two. As a first, it violates the transitivity of the evolution procedure. As a consequence, the comparison of different  evolution schemes is possible only  with a work of reverse engineering of equations which can  be also  un-precise.  Ultimately, this  destroys also the concept of universality of the non-perturbative functions. Secondly, it is difficult (but not impossible) to trace internal inconsistencies of the phenomenological applications.  An efficient realization of the perturbative  part of the TMD is fundamental to provide a correct interpretation of the  QCD non-perturbative information.

The dissection  of the cross-section using the factorization theorem and the asymptotic limit of Operator Product Expansion (OPE) puts into evidence several important constituents of the TMD formalism beyond the evolution of TMD distributions,  such as their asymptotic matching onto collinear functions, etc. Every step of this theoretical process is accompanied by the  introduction of specific matching scales that control the goodness of the factorization/expansion. Traditionally, one sets up the scales to minimize the impact of individual logarithms in accordance to a classical one-dimensional evolution picture. However, the double-scale evolution grants an unprecedented freedom to set up the scales, if all scales are fixed coordinately. In this work we describe the fundamental origin of this freedom, and give the non-perturbative definition of the $\zeta$-prescription, which is a selection of scales that completely eliminate double logarithm contribution. Additionally, in the $\zeta$-prescription one completely disentangle the notion of the modeling of TMD distribution from the influence of TMD evolution. Altogether, the set of prescriptions that we  propose leads us to  obtain what we think is an  {\bf optimal TMD distribution}.

The  TMD evolution is  also affected by an additional complication  coming from the fact that it is partially non-perturbative. In other  words,  we need to match the perturbative and non-perturbative  parts of the TMD evolution. This issue  has been discussed in several works in the literature (see e.g.~\cite{Collins:2011zzd,Aybat:2011zv,Echevarria:2012pw}). The renormalon nature of this behavior has been known for many years in~\cite{Korchemsky:1994is} and the object of explicit calculations~\cite{Scimemi:2016ffw}. Thus, one should not be surprised that the non-perturbative effects included in this part of evolution strongly depend on prescriptions used to solve the solution ambiguity. 

The article is structured  as in the following. In the first part, given in the sec.~\ref{sec:1part}, we present the elementary theory of TMD evolution, stressing its two dimensional nature.  In order to emphasize it, we introduce the vector notation and the concept of the \lq\lq{}evolution field\rq\rq{}, which allows multiple analogies to mathematical physics. We explicitly demonstrate the freedoms granted by the two dimensional nature of the TMD evolution, such as the freedom in the selection of the scales, contours of integrations, etc., which has not been used so far. We also discuss the  structure of singularities of the evolution field, that gives a new point of view of some well-known concepts.

In the second part of the paper we discuss the mathematical aspects of TMD evolution in the truncated perturbation theory (sec.~\ref{sec:truncPT}) and show that it leads to an ambigous  TMD evolution. In sec.~\ref{sec:restoreP} we discuss the opportunities to fix the ambiguity. In particular, we demonstrate that the traditional method of "resummed"  rapidity anomalous dimension (or Sudakov exponentiation) does not entirely solve the problem, but only reduce the uncertainties. From our side we suggest an alternative method to fix the evolution by "improving" the ultraviolet anomalous dimension. The suggested method is simple, obeys all expected demands, and it is easily generalizable to any model of non-perturbative evolution.

In the third part of the paper, given in sec.~\ref{sec:zeta-prescription}, we discuss the role of scale choices in the definition of TMD distributions, and introduce the concept of $\zeta$-prescription. We show that the $\zeta$-prescription is a general feature of double-scale evolution. This feature has been completely overlooked in the applications. The particular realization of $\zeta$-prescription that is characterized by the absence of any restriction on the model for TMD distribution defines the  optimal TMD distribution. As the standard selection of scales gives no benefits, we suggest here the optimal TMD distribution as a universal object for phenomenological studies.

Finally we collect the formulas needed for a generic TMD cross-section of a  Drell-Yan or  SIDIS process and we  resume our findings. Using these cases, we also recall the perturbative series that enters the cross-sections and systematize the sources of perturbative uncertainties, checking the  variation of all relevant scales in several examples. We observe directly that the solution that we propose, with the implementation of the optimal TMD, reshuffles the distribution of theoretical errors and, globally, it provides  a better control of theoretical uncertainties. 

\section{General structure of TMD evolution}
\label{sec:1part}

The purpose of this section is to provide the basic concepts and notation  for the TMD distributions and their evolution equations. We also introduce a convenient vector notation, which makes transparent  some properties of the evolution of TMD distributions which should  taken into account carefully. Everywhere in this section, we consider every perturbative series as un-truncated, so their properties can be  easily established.  Many results of the section could be translated to the cases of other double-scale functions.

\subsection{Definition of anomalous dimensions}

The evolution of the TMD distributions (or TMD evolution for simplicity) is given by the following pair of equations
\begin{eqnarray}\label{def:TMD_ev_UV}
\mu^2 \frac{d}{d\mu^2} F_{f\ot h}(x,b;\mu,\zeta)&=&\frac{\gamma^f_F(\mu,\zeta)}{2}F_{f\ot h}(x,b;\mu,\zeta),
\\\label{def:TMD_ev_RAP}
\zeta\frac{d}{d\zeta}F_{f \ot h}(x,b;\mu,\zeta)&=& -\mathcal{D}^f(\mu,b)F_{f\ot h}(x,b;\mu,\zeta),
\end{eqnarray}
where $F_{f\ot h}$ is the TMD parton distribution function (TMDPDF) of the parton $f$ in hadron $h$. The argument $x$ is the usual Bjorken variable, and $b$ is the transverse distance. The evolution equations for TMD fragmentation functions (TMDFF, and symbolically $D_{f\to h}$) have the same form with the replacement of $F_{f\ot h}$ by $D_{f\to h}$. For the exact field theoretical definition of TMD distributions see e.g.~\cite{Echevarria:2016scs}. The equation (\ref{def:TMD_ev_UV}) is a standard renormalization group equation, which comes from the renormalization of the ultraviolet divergences.  {\bf The function $\gamma_F(\mu,\zeta)$ is called the TMD anomalous dimension} and contains both single and double logarithms (see e.g. definition in~\cite{Echevarria:2016scs} and eq.~(\ref{th:gammaV})). The equation (\ref{def:TMD_ev_RAP}) results from the factorization of rapidity divergences (for the detailed description see e.g. ~\cite{Echevarria:2015usa,Vladimirov:2016qkd,Vladimirov:2017ksc}). {\bf The function $\mathcal{D}(\mu,b)$ is called the rapidity anomalous dimension}. TMD and rapidity anomalous dimensions have not unified notation in the literature. The notations $\gamma_F$ and $\mathcal{D}$, used in this article, were suggested in~\cite{Echevarria:2012pw}. For convenience we list some popular notations and their relation to our notation in the table~\ref{tab:ADs}.

\begin{table}[b]
\renewcommand{\arraystretch}{1.5}
\begin{tabular}{l || c | c | c | c |  c}
 & ~~\parbox[b]{1.7cm}{rapidity evolution scale}~~ & ~~\parbox[b]{1.7cm}{TMD anomalous dimension}~~ & ~~\parbox[b]{1.7cm}{cusp anomalous dimension}~~ &   & \parbox[b]{1.7cm}{rapidity anomalous dimension} 
 \\ \hline\hline
here \& \cite{Echevarria:2012pw,Echevarria:2015usa,Scimemi:2017etj} & $\zeta$ & $\gamma_F$ & $\Gamma$ & $\gamma_V$ &$\mathcal{D}$ 
\\\hline
\cite{Collins:2011zzd,Aybat:2011zv} & $\zeta$ & $\gamma_F ~(=\gamma_D)$& $\frac{1}{2}\gamma_K$ & ~~$-\gamma_F(g(\mu);1)$~~ & $-\frac{1}{2}\tilde K$
\\\hline
\cite{Becher:2010tm,Becher:2011xn,Gehrmann:2014yya} & -- & -- & $\Gamma_{cusp}$ & $2\gamma^q$ & $\frac{1}{2}F_{f\bar f}$
\\\hline
\cite{Chiu:2012ir} & $\nu^2$ & $\gamma_\mu^{f_\perp}$ & $\Gamma_{cusp}$ & -- & $-\frac{1}{2}\gamma_\nu^{f_\perp}$
\end{tabular}
\caption{\label{tab:ADs} Correspondence of notation for TMD anomalous dimensions used here to some other popular notations.}
\end{table}

Starting from the  definition of TMD operators, whose matrix elements give the TMD distributions, some properties  of the evolution have already been established  in the past. The evolution equations are independent of quantum numbers of the hadrons which enter in the TMD distributions, because they are properties of the TMD operators. Moreover, they do not depend  on the polarization of partons~\cite{Collins:2011zzd,GarciaEchevarria:2011rb,Echevarria:2016scs,Gutierrez-Reyes:2017glx} and they are the same for TMDPDF and TMDFF (at least at the two-loop order, see \cite{Echevarria:2016scs}). Altogether, these properties describe the \textit{universality} of TMD evolution. The only important quantum number for TMDs is the color representation the initiating parton, which is tied to the parton flavor, namely, quark (fundamental representation) or gluon (adjoint representation). However,  as the TMD evolution does not mix the flavors and  for simplicity of  notation, we omit the flavor index $f$ in  most of the article. The restoration of the flavor index is straightforward.

The equation (\ref{def:TMD_ev_UV})-(\ref{def:TMD_ev_RAP}) are  coupled, due to the fact that the ultraviolet divergences of the TMD operator partially overlap with the rapidity divergences. As a result, the anomalous dimensions  of the two scales are correlated. The mutual dependence can be worked out  explicitly (see e.g.\cite{Collins:2011zzd,Chiu:2012ir,Vladimirov:2017ksc}), 
\begin{eqnarray}\label{th:dGammaF=G}
\zeta \frac{d}{d\zeta}\gamma_F(\mu,\zeta)=-\Gamma(\mu),
\\\label{th:dDD=G}
\mu \frac{d}{d\mu}\mathcal{D}(\mu,b)=\Gamma(\mu),
\end{eqnarray}
where $\Gamma$ is the (light-like) cusp anomalous dimension. The equation (\ref{th:dGammaF=G}) entirely fixes the logarithm dependence of the TMD anomalous dimension, which reads
\begin{eqnarray}\label{th:gammaV}
\gamma_F(\mu,\zeta)=\Gamma(\mu) \ln\(\frac{\mu^2}{\zeta}\)-\gamma_V(\mu).
\end{eqnarray}
The anomalous dimension $\gamma_V$ refers to the finite part of the renormalization of the vector form factor. In contrast, the equation (\ref{th:dDD=G}) cannot fix the logarithmic part of $\mathcal{D}$ entirely, but only order by order in  perturbation theory, because the parameter $\mu$ is also responsible for the running of the coupling constant. It has been shown~\cite{Scimemi:2016ffw} that the perturbative series for $\mathcal{D}$ is asymptotical and it has a renormalon pole, whose contribution is significant at large-$b$. Therefore, the rapidity anomalous dimension $\mathcal{D}$ is generically a non-perturbative function, which admits a perturbative expansion only for small values of the parameter $b$. On the other side, in conformal field theory, where the coupling constant is independent on $\mu$, the rapidity anomalous dimension is linear in logarithms of $\mu_b$ and coincides with the soft anomalous dimension~\cite{Vladimirov:2016dll,Vladimirov:2017ksc}.

\subsection{General properties  of the TMD evolution factor}

The solution of eq.~(\ref{def:TMD_ev_UV})-(\ref{def:TMD_ev_RAP})  can be written as
\begin{eqnarray}\label{th:TMD_evol}
F(x,b;\mu_f,\zeta_f)=R[b;(\mu_f,\zeta_f)\to (\mu_i,\zeta_i)]F(x,b;\mu_i,\zeta_i),
\end{eqnarray}
where {\bf $R$ is the TMD evolution factor}. The uniqueness of solution for the system (\ref{def:TMD_ev_UV})-(\ref{def:TMD_ev_RAP}) is guaranteed by \textit{the integrability condition}
\begin{eqnarray}\label{th:consitency}
\zeta \frac{d}{d\zeta}\gamma_F(\mu,\zeta)=-\mu \frac{d}{d\mu}\mathcal{D}(\mu,b),
\end{eqnarray}
which obviously follows from the equations (\ref{th:dGammaF=G}) and (\ref{th:dDD=G}). 

The general form of the evolution factor is
\begin{eqnarray}\label{th:TMD_R}
R[b;(\mu_f,\zeta_f)\to (\mu_i,\zeta_i)]=\exp\[\int_P \(\gamma_F(\mu,\zeta)\frac{d\mu}{\mu} -\mathcal{D}(\mu,b)\frac{d\zeta}{\zeta}\)\] ,
\end{eqnarray}
where $(\mu_f,\zeta_f)$ and $(\mu_i,\zeta_i)$ refer respectively to a final  and initial set of scales. Here, the $\int_P$ denotes the line integral along the path $P$ in the $(\mu,\zeta)$-plane from the point $(\mu_f,\zeta_f)$ to the point $(\mu_i,\zeta_i)$. The integration can be done on an arbitrary path $P$, and the solution is independent on it, thanks to the integrability condition eq.~(\ref{th:consitency}). 

The TMD evolution factor $R$ obeys the transitivity relation
\begin{eqnarray}\label{th:transitivity}
R[b;(\mu_1,\zeta_1)\to (\mu_2,\zeta_2)]=R[b;(\mu_1,\zeta_1)\to (\mu_3,\zeta_3)]R[b;(\mu_3,\zeta_3)\to (\mu_2,\zeta_2)],
\end{eqnarray}
where $(\mu_3,\zeta_3)$ is arbitrary point in $(\mu,\zeta)$-plane and the  point inversion property
\begin{eqnarray}\label{th:inversion}
R[b;(\mu_1,\zeta_1)\to (\mu_2,\zeta_2)]=R^{-1}[b;(\mu_2,\zeta_2)\to (\mu_1,\zeta_1)].
\end{eqnarray}
These equations are the cornerstones of the evolution mechanism, since they allow an universal definition of the non-perturbative distributions and the comparison of different experiments. 

In practical applications  one then has to make a choice for the initial and final scales. For  the final scales
the typical choice is the hard scale appearing in the the process, $Q$ (see also sec.~\ref{subsec:tradition}). So,
\begin{eqnarray}\label{th:muzetaf}
(\mu_f,\zeta_f)=(Q,Q^2),
\end{eqnarray}
and of course $Q\gg \Lambda$.  

The initial scale  $(\mu_i,\zeta_i)$, instead, is the scale where the non-perturbative input for TMD distributions is inserted. This non-perturbative input is usually provided by models, and it is not a subject of TMD factorization. A typical model for TMD distributions incorporates the small-$b$ operator product expansion (OPE), which matches the TMD distributions with integrated distributions and improves the prediction power for high-energy experiments. In this case, the model for TMD distribution has the form
\begin{eqnarray}\label{th:OPE}
F(x,b;\mu_i,\zeta_i)\sim\sum_{n}C_n(x,\mathbf{L}_{\mu_i};\mu_i,\zeta_i)\otimes f_n(x,\mu_i),
\end{eqnarray}
where $C$ is the Wilson coefficient function and $\otimes$ is a convolution in  the Bjorken variable $x$. Here and in the following we use the notation
\begin{eqnarray}\label{def:Lmu}
\mathbf{L}_X=\ln\(\frac{X^2 b^2}{4 e^{-2\gamma_E}}\).
\end{eqnarray}
The request for minimization of the logarithmic contributions in the coefficient function in eq.~(\ref{th:OPE}) dictates the choice of initial scale $\mu_i\sim b^{-1}$. Let us emphasize here that the parameter $\zeta$ remains unrestricted. Often (see e.g.~\cite{Aybat:2011zv,Collins:2011zzd}), one sets $\zeta_i=\mu_i^2$. This choice is naively justified by the elimination of $\ln \mu_i^2/\zeta_i$ from coefficient function, but actually is not the ideal one. In sec.~\ref{sec:truncPT} and sec.~\ref{sec:zeta-prescription}, we critically analyze these common choices, and suggest another selection of scales that guarantees the minimization of the logarithmic contribution in eq.~(\ref{th:OPE}) on the whole range of $b$.

\begin{figure}[t]
\centering
\includegraphics[width=0.4\textwidth]{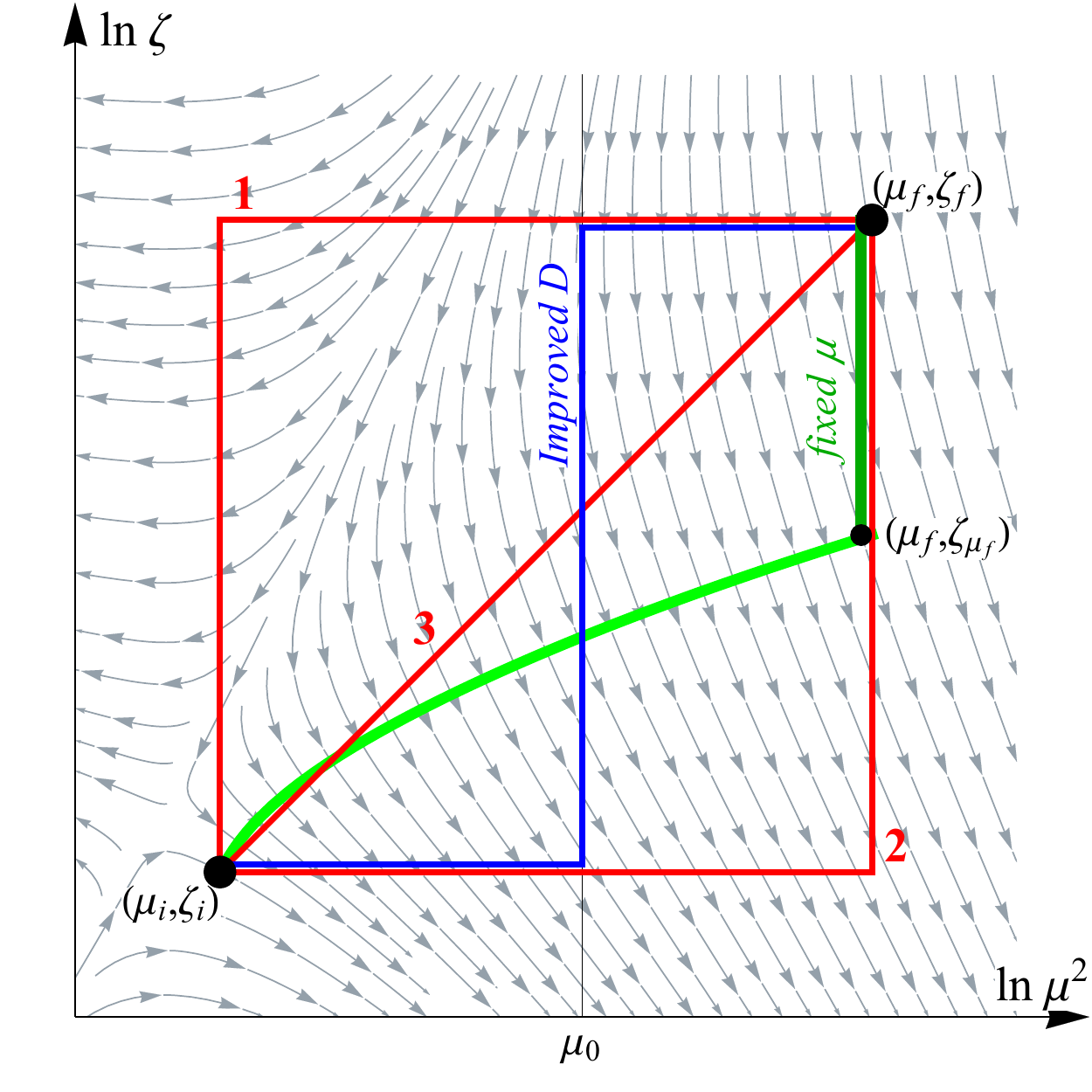}
\caption{Illustration of evolution paths corresponding to different solutions. Red lines show the solutions 1, 2, and 3, defined in eqns.~(\ref{th:Rpath1}), (\ref{th:Rpath2}), and (\ref{th:Rpath1}), correspondingly. The blue line shows the path of the improved $\mathcal{D}$ solution (\ref{th:RimprovedD}) with the normalization point $\mu_0$. Green line shows the path of the fixed-$\mu$ solution, (\ref{th:R-fixedmu}). The light-green curve shows the null-evolution curve which passes though the point $(\mu_i,\zeta_i)$. The evolution along light-green curve is absent.} \label{fig:solutions}
\end{figure}

Another important point in the implementation of the TMD evolution factor $R$, is represented by the integration path. The TMD evolution factor $R$ is path independent, however in practice, one has to provide a choice. The two simplest choices of integration paths are the  combinations of straight segments as
\begin{eqnarray*}
\text{path 1}&:& (\mu_f,\zeta_f)\to (\mu_i,\zeta_f)\to (\mu_i,\zeta_i),
\\
\text{path 2}&:& (\mu_f,\zeta_f)\to (\mu_f,\zeta_i)\to (\mu_i,\zeta_i).
\end{eqnarray*}
In the first path  the evolution is along $\mu$ first and then along $\zeta$, while in the second path  the evolution is along $\zeta$ first  and subsequently along $\mu$. In the $(\mu,\zeta)$-plane these paths form a rectangle, see fig.~\ref{fig:solutions}. We call the solutions corresponding to these paths  as solutions 1 and 2, for simplicity. Their explicit forms are
\begin{eqnarray}\label{th:Rpath1}
\text{solution 1}: \qquad \ln R[b;(\mu_f,\zeta_f)\xrightarrow{1} (\mu_i,\zeta_i)]&=&\int_{\mu_i}^{\mu_f}\frac{d\mu}{\mu}\gamma_F(\mu,\zeta_f)-\mathcal{D}(\mu_i,b)\ln\(\frac{\zeta_f}{\zeta_i}\),
\\\label{th:Rpath2}
\text{solution 2}: \qquad \ln R[b;(\mu_f,\zeta_f)\xrightarrow{2} (\mu_i,\zeta_i)]&=&\int_{\mu_i}^{\mu_f}\frac{d\mu}{\mu}\gamma_F(\mu,\zeta_i)-\mathcal{D}(\mu_f,b)\ln\(\frac{\zeta_f}{\zeta_i}\).
\end{eqnarray}
The solution 1 is practically the only one used in the literature, since it has the form of the resummed Sudakov exponent, see e.g.~\cite{Dokshitzer:1978hw,Collins:1981uk,Collins:1981va}.

In the next section we discuss the effects of violation of path-independence. So that the solutions 1 and 2 can serve as natural extreme cases. For comparison, we also introduce an intermediate solution whose path has the form of a straight line between points $(\mu_f,\zeta_f)$ and $(\mu_i,\zeta_i)$. We call it the solution 3. Its explicit form reads
\begin{eqnarray}\label{th:Rpath3}
\text{solution 3}:\qquad &&\ln R[b;(\mu_f,\zeta_f)\xrightarrow{3} (\mu_i,\zeta_i)]=
\\\nn&& \qquad \int_0^1 \(\gamma_F(\mu(t),\zeta(t))\frac{\mu_f-\mu_i}{\mu(t)} -\mathcal{D}(\mu(t),b)\frac{\zeta_f-\zeta_i}{\zeta(t)}\)dt,
\end{eqnarray}
where $t$ parameterizes the path of integration, $\mu(t)=(\mu_f-\mu_i)t+\mu_i$ and $\zeta(t)=(\zeta_f-\zeta_i)t+\zeta_i$.

\subsection{Two-dimensional notation and the scalar potential for TMD evolution}

The TMD evolution is  naturally formulated in the terms of two-dimensional vectors and fields. In this section, we introduce the vector notation and rewrite the main equations of the previous sections. By the bold font we designate the two-dimensional vectors.

Let us introduce the convenient two-dimensional variable which treats scales $\mu$ and $\zeta$ equally,
\begin{eqnarray}\label{def:nu}
\vec \nu=(\ln \(\frac{\mu^2}{\text{1 GeV}^2}\),\ln \(\frac{\zeta}{\text{1 GeV}^2}\)).
\end{eqnarray}
Here the notation 1 GeV$^2$ is set to indicate the unit transformation from the dimensional parameters $\mu$ and $\zeta$ to dimensionless $\vec \nu$. The particular value of normalization plays no role in the following discussion, but could be easily reconstructed if necessary.  We also define the standard vector differential operations in the plane $\vec \nu$, namely, the gradient and the curl
\begin{eqnarray}\label{def:grad+curl}
\vec \nabla=\frac{d}{d\vec \nu}=(\mu^2\frac{d}{d\mu^2},\zeta\frac{d}{d\zeta}),\qquad \textbf{curl}=(-\zeta\frac{d}{d\zeta},\mu^2\frac{d}{d\mu^2}).
\end{eqnarray}

The TMD evolution is defined by the anomalous dimension which form the vector field $\mathbf{E}(\vec \nu,b)$. Explicitly, it is defined as
\begin{eqnarray}
\mathbf{E}(\vec \nu,b)=(\frac{\gamma_F(\vec \nu)}{2},-\mathcal{D}(\vec \nu,b)).
\end{eqnarray}
Here and in the following, we use the vectors $\vec \nu$ as the argument of the anomalous dimensions for brevity, keeping in  mind that $\mathcal{D}(\vec \nu,b)=\mathcal{D}(\mu,b)$, $\gamma_F(\vec \nu)=\gamma_F(\mu,\zeta)$, etc. In other words, the anomalous dimensions are to be evaluated on the corresponding values of $\mu$ and $\zeta$ defined by value of $\vec \nu$ in eq.~(\ref{def:nu}). The TMD evolution equations (\ref{def:TMD_ev_UV},~\ref{def:TMD_ev_RAP}) in this notation have the form
\begin{eqnarray}\label{def:TMD_ev_nu}
\vec \nabla F(x,b;\vec \nu)=\mathbf{E}(\vec \nu,b)F(x,b;\vec \nu),
\end{eqnarray}
and thus the vector field $\mathbf{E}$ has the meaning of the evolution flow field. Correspondingly, the TMD evolution factor (\ref{th:TMD_R}) reads
\begin{eqnarray}\label{th:R_in2D_notation}
\ln R[b,\vec \nu_f\to \vec \nu_i]=\int_P \mathbf{E}\cdot d\vec \nu.
\end{eqnarray}
Written in such form the TMD evolution suggests multiple analogies with different branches of physics.

Individually, the equations (\ref{th:dGammaF=G},~\ref{th:dDD=G}) do not imply any special geometrical meaning. In contrast, the integrability condition in eq.~(\ref{th:consitency}) that can be seen as a consequence of eqns.~(\ref{th:dGammaF=G},~\ref{th:dDD=G}), has a deep meaning and it is equivalent to the statement that the evolution flow is \textit{irrotational},
\begin{eqnarray}\label{def:irrotational}
\vec \nabla \times \mathbf{E}=0.
\end{eqnarray}
The irrotational vector fields are also known as \textit{conservative} fields, and they can be presented as a gradient of a \textit{scalar potential},
\begin{eqnarray}\label{def:scalarP}
\mathbf{E}(\vec \nu,b)=\vec \nabla U(\vec \nu,b),
\end{eqnarray}
i.e.  \textbf{$U$ is the scalar potential for TMD evolution}. According to the gradient theorem any line integral of the field $\mathbf{E}$ is path-independent and equals to the difference of values of potential at end-points. Therefore, the solution in eq.~(\ref{def:TMD_ev_nu}) can be presented as
\begin{eqnarray}\label{th:potential-solution}
\ln R[b;\vec \nu_f\to\vec \nu_i]=U(\vec \nu_f,b)-U(\vec \nu_i,b).
\end{eqnarray}
In this form the evolution kernel is explicitly path-independent and obeys the transitivity property in eq.~(\ref{th:transitivity}). The explicit form of the scalar potential can be found by integrating   eq.~(\ref{def:scalarP}), namely
\begin{eqnarray}\label{th:potential_explicit}
U(\vec \nu,b)=\int^{\nu_{1}} \frac{\Gamma(s)s-\gamma_V(s)}{2}ds-\mathcal{D}(\vec \nu,b)\nu_2+\const(b),
\end{eqnarray}
where $\nu_{1,2}$ are the components of the vector $\vec \nu$ in eq.~(\ref{def:nu}), and the last term is an arbitrary $b$-dependent function.

\subsection{Singularities on evolution plane}
\label{sec:singularitiesOfEvPlane}

\begin{figure}[t]
\centering
\includegraphics[width=0.32\textwidth]{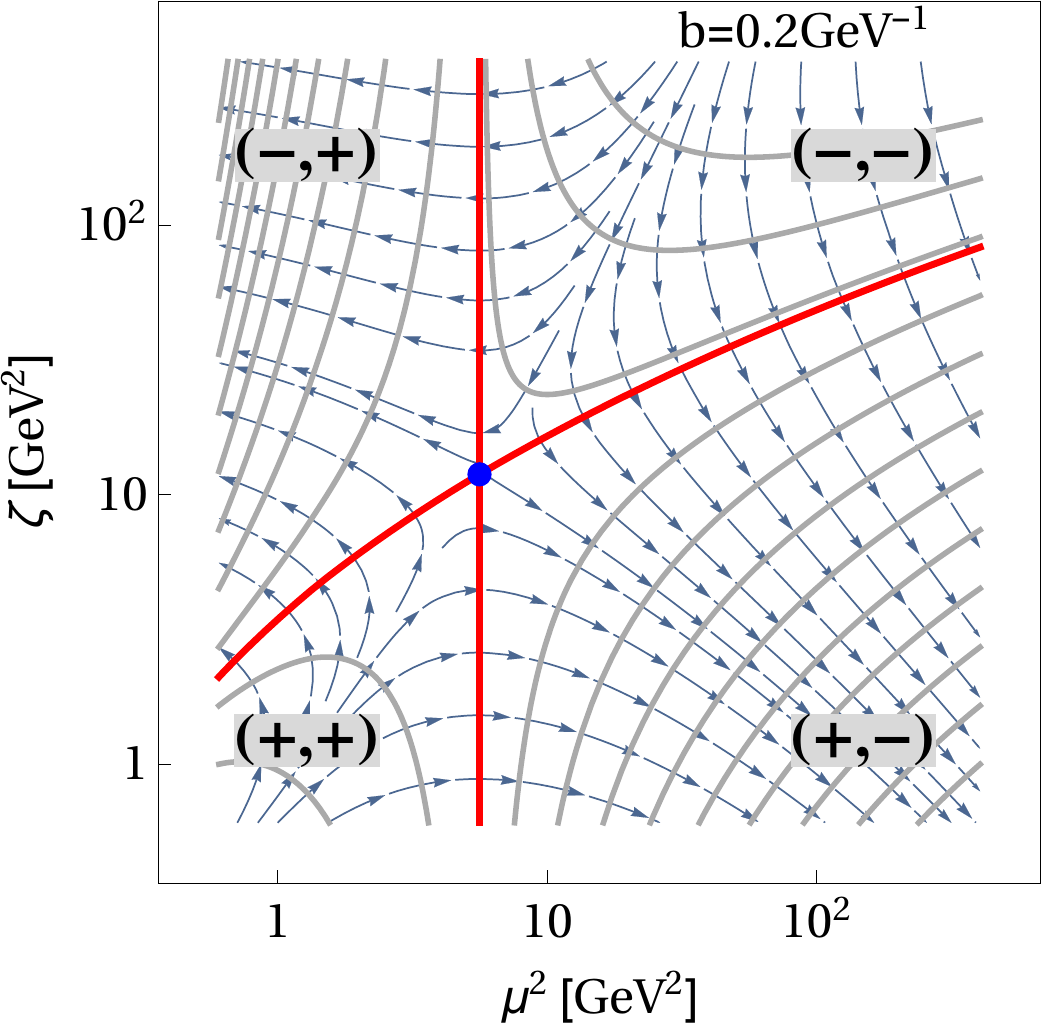}
\includegraphics[width=0.32\textwidth]{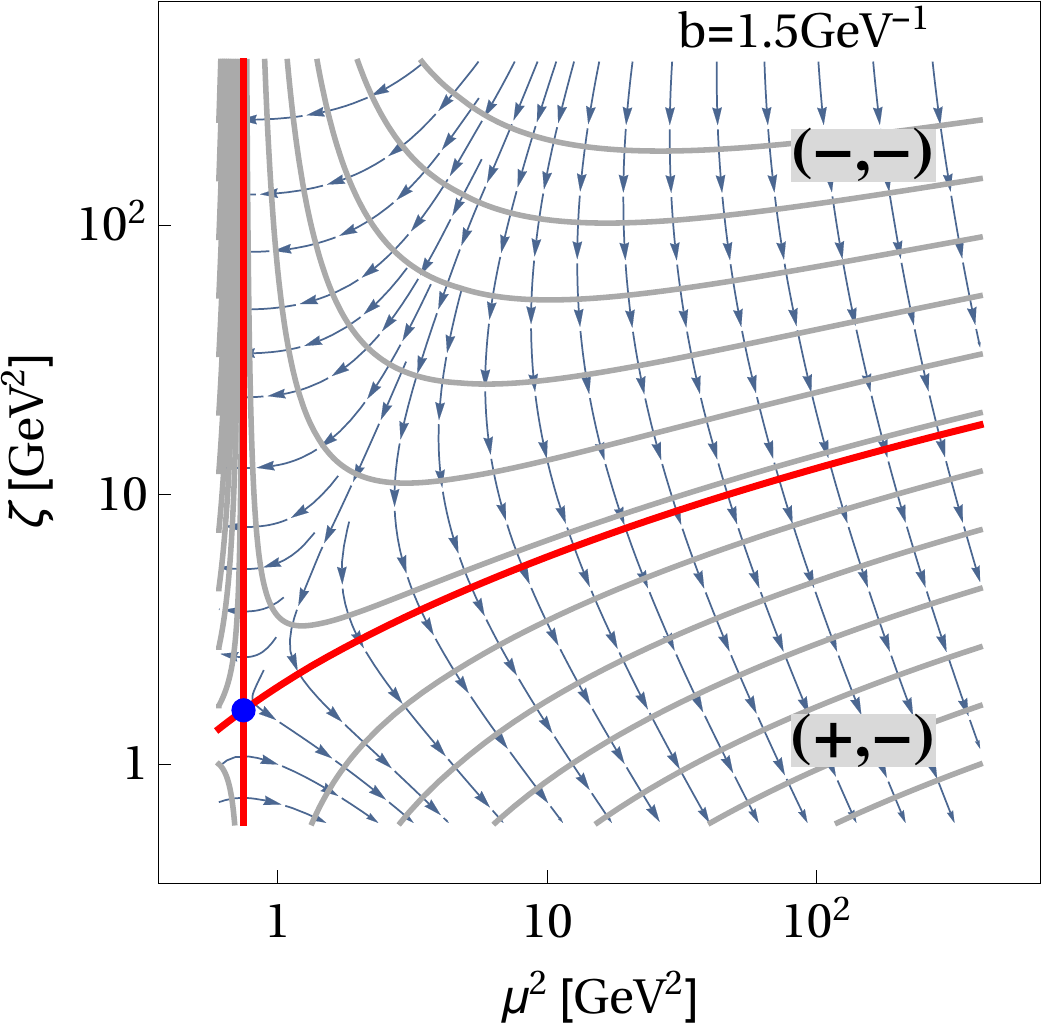}
\includegraphics[width=0.32\textwidth]{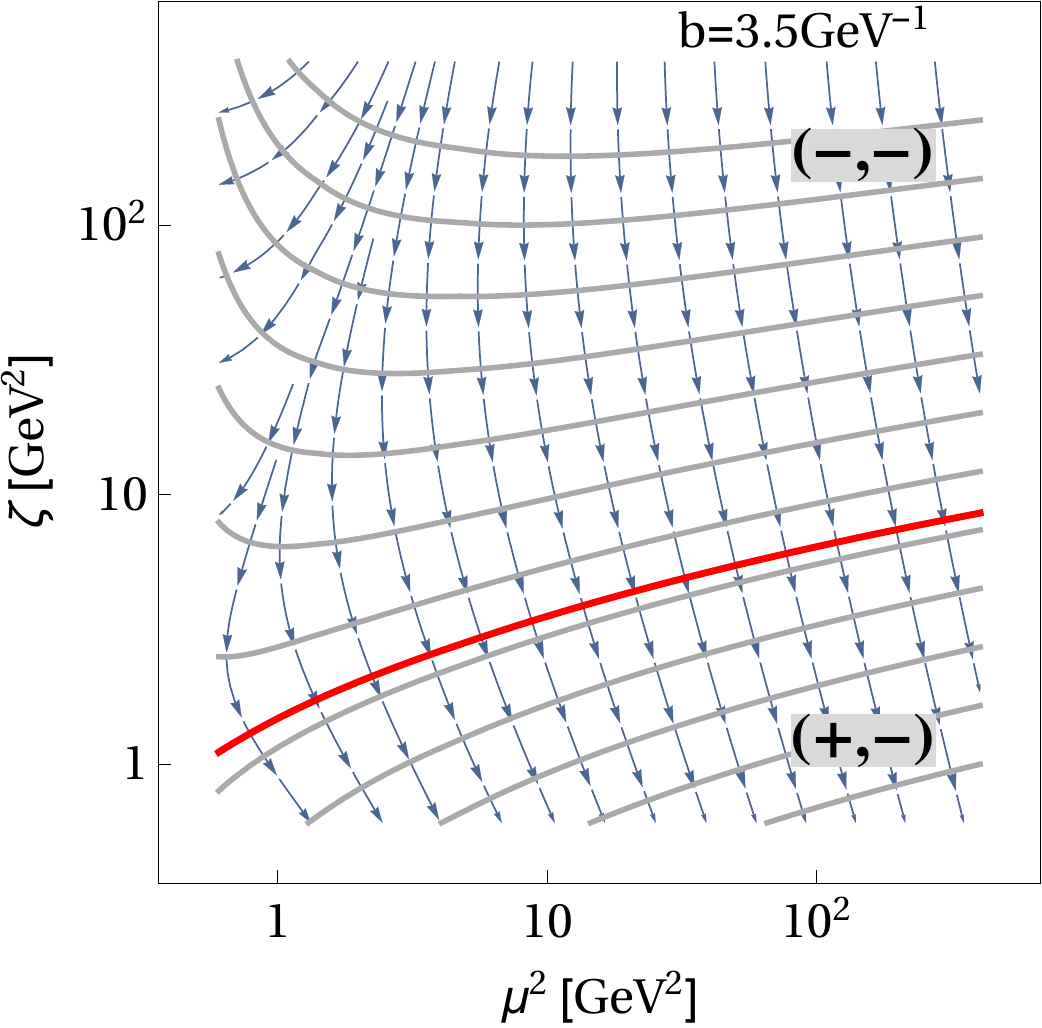}
\caption{The illustration of evolution flow field $\mathbf{E}$ at different values of $b$. The blue point is the stable point. Gray curves are the equipotential lines (null-evolution curves). Red curves are special null-evolution curves. Special curves split the plane into quadrants with preserved sign of field components, which is shown by bold font. At large $b$ the stable point moves to the values of smaller $\mu$, and crosses the Landau pole. The line of Landau pole in not presented and it is located at smaller values of $\mu$.} \label{fig:EvolutionPlane}
\end{figure}

The evolution flow and the scalar potential have a non-trivial structure which is discussed in the present and in the following sections. The graphical representation of the evolution flow is shown in fig.~\ref{fig:EvolutionPlane}. 

It is of great importance to classify the singularities of the scalar potential and the evolution flow. In particular we are interested in the singularities that are located at finite values of parameters. There are two of them. First, there is the line $\mu=\Lambda$ (where $\Lambda$ is the position of the Landau pole) at which both components of $\mathbf{E}$ turn to infinity. On top of  this line, and for smaller $\mu$ the scalar potential is undefined. In fig.~\ref{fig:EvolutionPlane} this line is not shown and it is located on the left  side of the plotted region. Second, there is a saddle point where both components of $\mathbf{E}$ turn to zero. In fig.~\ref{fig:EvolutionPlane} the saddle point is depicted by a blue dot. The position of the saddle point is dictated by the equation
\begin{eqnarray}
\mathbf{E}(\vec \nu_{\text{saddle}},b)=\vec 0.
\end{eqnarray}
In the standard notation this equation reads
\begin{eqnarray}\label{def:saddle_values}
\mathcal{D}(\mu_{\text{saddle}},b)=0,\qquad \zeta_{\text{saddle}}=\mu^2_{\text{saddle}}\exp\(-\frac{\gamma_V(\mu_{\text{saddle}})}{\Gamma(\mu_{\text{saddle}})}\).
\end{eqnarray}
At one-loop these equations are functionally independent on $a_s(\mu)$ and the saddle point position can be found explicitly. This value can be used as a good approximation of saddle point position (here for the quark flavor)
\begin{eqnarray}\label{th:saddle_crude}
\mu_{\text{saddle}}\approx\frac{2e^{-\gamma_E}}{b},\qquad \zeta_{\text{saddle}}\approx \frac{4 e^{-2\gamma_E+\frac{3}{2}}}{b^2}.
\end{eqnarray}
The position of saddle point depends on the parameter $b$, see fig.~\ref{fig:saddle}. Generally, it moves to  larger values of $\mu$ and $\zeta$ for  smaller-$b$. In particular, at some (large) value $\bar b$ the saddle point crosses the Landau pole line and escapes the observable region. Using eq.~(\ref{th:saddle_crude}) we can estimate that $\bar b\approx 2 e^{-\gamma_E}/\Lambda \approx 4$GeV$^{-1}$.

\subsection{Null-evolution curves}
\label{sec:null-evolution}

The \textit{equipotential} curves play the special role. Along these curves the scalar potential for TMD evolution does not change its value, and consequently the TMD evolution is 1 (unity) between points laying on the same equipotential curve. For this reason {\bf the equipotential curves are also  \textit{null-evolution} curves}. 

Let us denote the equipotential curve which passes through the point $\vec \nu_B$ as $\vec\omega(t,\vec \nu_B,b)$. This curve is also a solution of 
\begin{eqnarray}\label{th:equipotential_eqn}
\frac{d\vec \omega}{dt} \cdot \vec \nabla U(\vec \omega,b)=0,
\end{eqnarray}
where $t$ parameterizes the curve $\vec \omega$. A convenient parameterization of equipotential curve is
\begin{eqnarray}
\vec \omega(t,\vec \nu_B,b)=(t,\omega(t,\vec \nu_B,b)),
\end{eqnarray}
where we identify the first component of the vector $\vec \omega$ with the parametrization parameter. In this form  eq.~(\ref{th:equipotential_eqn}) turns into
\begin{eqnarray}\label{th:equipotential_eqn_simple}
\gamma_F(\vec \omega)-2\mathcal{D}(\vec \omega,b)\omega'(t)=0,
\end{eqnarray}
where we omit the arguments $\vec \nu_B$ and $b$ of the function $\omega$ for brevity. The solution of this equation reads
\begin{eqnarray}\label{th:equipotential_lines}
\omega(t,\vec \nu_B,b)=\omega_B(b) e^{-\int_{t_0}^t \frac{\Gamma(r)}{2\mathcal{D}(r,b)}dr}+\int_{t_B}^t  e^{-\int_{s}^t \frac{\Gamma(r)}{2\mathcal{D}(r,b)}dr}\frac{\Gamma(s)s-\gamma_V(s)}{2\mathcal{D}(s,b)}ds,
\end{eqnarray}
where $t_B=(\nu_B)_1$ and $\omega_B=(\nu_B)_2$ are the components of the boundary condition $\vec \nu_B$. Using the connection of the derivative of rapidity anomalous dimension to cusp anomalous dimension (\ref{th:consitency}) we simplify the solution (\ref{th:equipotential_lines}) and obtain
\begin{eqnarray}\label{th:equipotential_lines2}
\omega(t,\vec \nu_B,b)=\frac{\omega_B(b)\mathcal{D}(t_0,b)+\int_{t_B}^t\frac{\Gamma(s)s-\gamma_V(s)}{2}ds}{\mathcal{D}(t,b)}.
\end{eqnarray}
This expression can be also obtained using the definition of equipotential curve as $U(\vec \omega)=U(\vec \nu_B)$, and the fact that the scalar potential in eq.~(\ref{th:potential_explicit}) is linear in $\nu_2$. 

Note, that there is an additional equipotential curve that is not included in the solution (\ref{th:equipotential_lines}). It is the line $\mu=\mu_{\text{saddle}}$. In  eq.~(\ref{th:equipotential_lines}) this line is singular.

The equipotential curves in eq.~(\ref{th:equipotential_lines}) do not intersect with each other with a single exception: the line $\mu=\mu_{\text{saddle}}$, and the line defined by eq.~(\ref{th:equipotential_lines}) with $\vec \nu_B=\vec \nu_{\text{saddle}}$. These lines intersect at the saddle point. For their selected definition we call these curves as \textbf{special null-evolution curves}. Special null-evolution curves are shown in fig.~\ref{fig:EvolutionPlane} by red lines. The evolution plane is cut by the special equipotential lines into quadrants and in each quadrant the sign of the components of the evolution field $\mathbf{E}$ is preserved. In particular, both components of $\mathbf{E}$ are negative in the first quadrant.

The evolution along any null-evolution curve is absent. This property can be used to simplify the explicit expression for the evolution kernel in eq.~(\ref{th:TMD_R}). Using the transitivity property of  $R$, eq.~(\ref{th:transitivity}), the evolution path can be split into two segments one of which is along an null-evolution curve, i.e.
\begin{eqnarray}\label{th:fixedMu-path}
R[b;\vec \nu_f\to\vec \nu_i]=R[b;\vec \nu_f\to \vec \omega(\vec \nu_i,b)]R[b;\vec \omega(\vec \nu_i,b) \to \vec \nu_i]=R[b;\vec \nu_f\to \vec \omega(\vec \nu_i,b)],
\end{eqnarray}
since $R[\vec \omega(\vec \nu_i) \to \vec \nu_i]=1$ by definition. The point  $ \vec \omega(\vec \nu_i,b)$ on the null-evolution curve can be selected arbitrarily. 
Nevertheless it is convenient to use the point with $t=\ln\mu^2_f$  so that the path of evolution has only a single vertical segment, see  the green curve in fig.~\ref{fig:solutions}. We address to this particular path as to \textbf{the fixed-$\mu$ solution}. In the standard notation the evolution kernel along  the fixed-$\mu$ solution path reads
\begin{eqnarray}\label{th:R-fixedmu}
\text{fixed-$\mu$ solution}:\qquad &&\ln R[b;(\mu_f,\zeta_f)\rightarrow (\mu_i,\zeta_i)]=
-\mathcal{D}(\mu_f,b)\ln\(\frac{\zeta_f}{\zeta_{\mu_f}(\mu_i,\zeta_i)}\),
\end{eqnarray}
where $\zeta_{\mu_f}(\mu_i,\zeta_i)$ is the $\zeta$-value of the null-evolution curve that passes though the point $(\mu_i,\zeta_i)$, at $\mu=\mu_f$.

\section{Effects of truncation of perturbation theory}
\label{sec:truncPT}

\begin{figure}[t]
\centering
\includegraphics[width=0.32\textwidth]{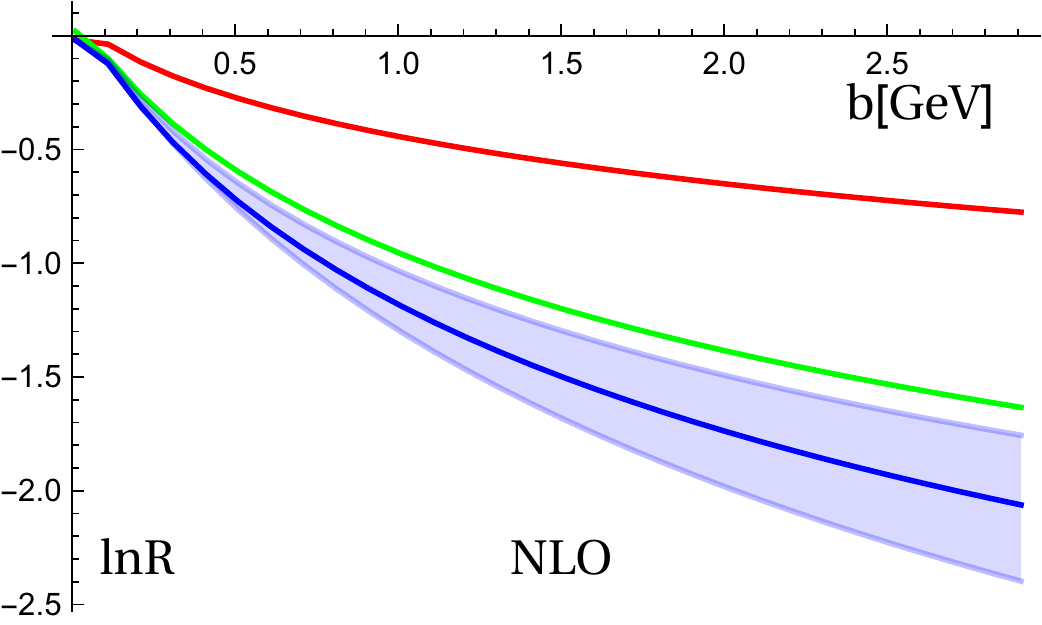}
\includegraphics[width=0.32\textwidth]{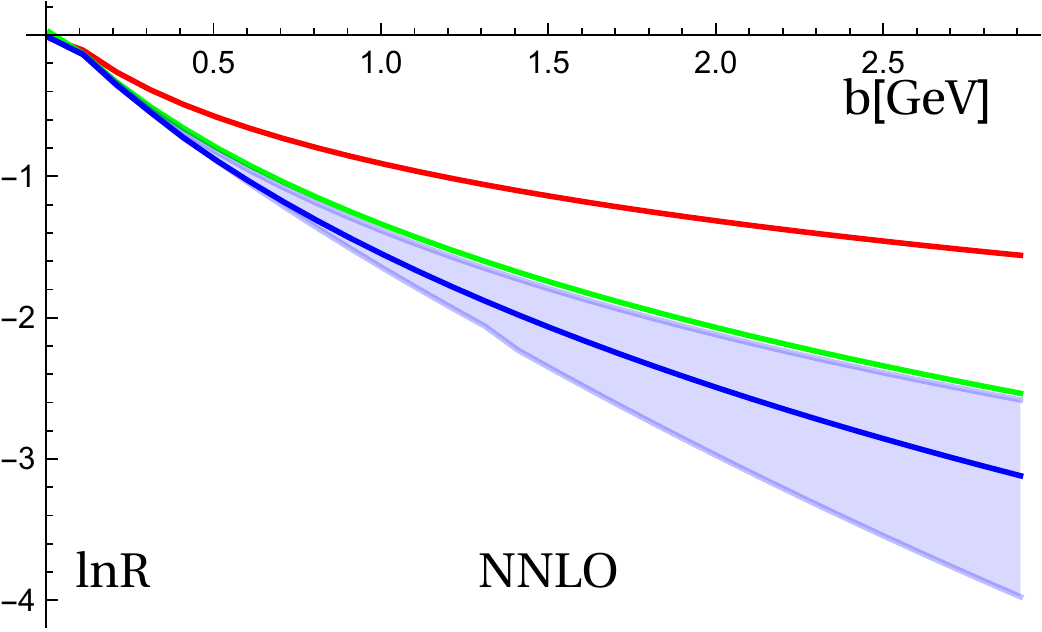}
\includegraphics[width=0.32\textwidth]{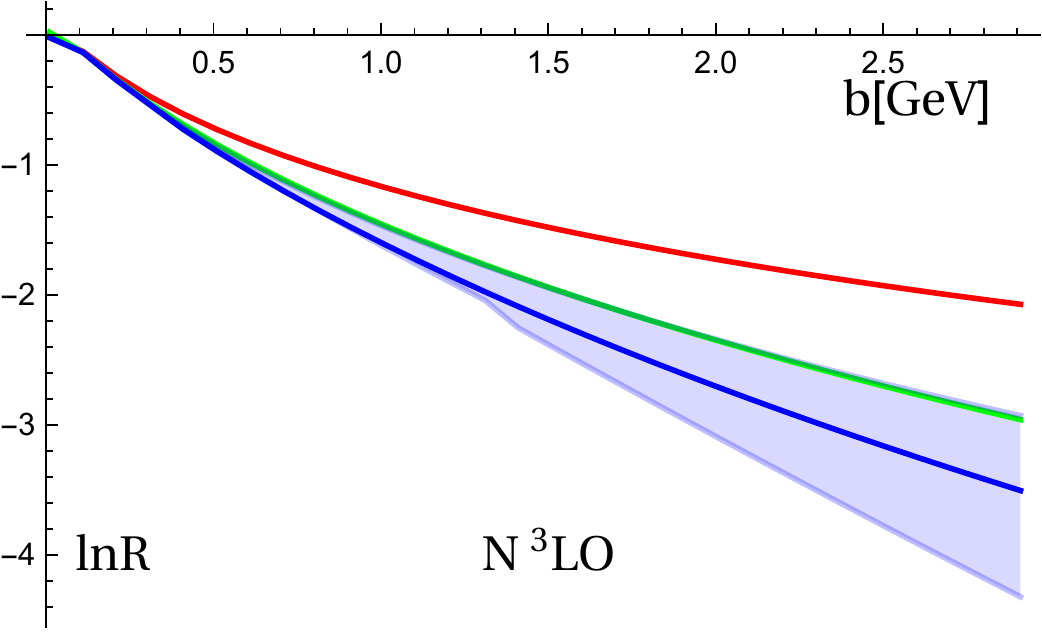}
\caption{Comparisons of different solution for $\ln R((M_Z,M_Z^2)\to(\mu_b,\mu_b^2))$ where $\mu_b=C_0/b+2$. The blue line is the solution 1. The red Line is the solution 2. The green line is the solution 3. The error band is obtained from the improved $\mathcal{D}$ solution at $\mu_0=\mu_i$ by variation of $\mu_0\in (0.5,2)\mu_i$. The blue line with error-band corresponds to the solution used in \cite{Scimemi:2016ffw}.} \label{fig:Comapre_lnR}
\end{figure}

The picture described above is idealistic. In real applications one operates with only a few terms of the perturbative series for the anomalous dimensions. Nowadays, these anomalous dimensions are known up to three-loop order (i.e. including term $a_s^3$ or up to NNLO), see~\cite{Moch:2004pa,Moch:2005id,Gehrmann:2010ue,Li:2016ctv,Vladimirov:2016dll}. In figs.~\ref{fig:Comapre_lnR} we show the function $R$ for different orders of perturbation theory and for different explicit path solutions given in eq.~(\ref{th:Rpath1},~\ref{th:Rpath2}) and (\ref{th:Rpath3}). The final point of the evolution is set to $Q=M_Z=91$GeV, which corresponds to the Z-boson production threshold. The initial point for the evolution has been set to $(\mu_i,\zeta_i)=(\mu_b,\mu_b^2)$ with  $\mu_b=2e^{-\gamma_E}/b+2$ GeV, as it has been used in~\cite{Scimemi:2017etj}. 

We observe that dissimilar realizations of $R$, which differ only by the integration path (and, in principle, are equivalent), produce enormous numerical differences.
Even at $b\sim 0.5$GeV$^{-1}$, which is still a  typical perturbative value (the strong coupling $a_s$ varies in the range $\sim 0.01 - 0.02$ within the evolution integral), the difference between solutions 1 and 2 is ($\sim 56$\%,$\sim 35$\%,$\sim 18$\%) at (LO, NLO, NNLO) respectively. The large spectrum in the values of the solution is clearly an effect of truncation of the perturbative series that is enhanced by the presence of logarithms in the rapidity anomalous dimension $\mathcal{D}$. In the case of solution 1 these logarithms are $\ln(\mu_f b)$, while for solution 2 these are $\ln(\mu_i b)$. This effect can be reduced by an appropriate resummation procedure. 

The path dependence of the solution leads to another potentially very dangerous problem, namely, the explicit violation of evolution transitivity and evolution inversion relations of  eq.~(\ref{th:transitivity},~\ref{th:inversion}). This effect  is  especially difficult to control. The path dependence prevents a clear direct comparison of fits when they are obtained with evolutions over different paths (which is practically always the case). Additionally,  the path dependence  makes more evident  that the shape of non-perturbative modifications of the rapidity anomalous dimension $\mathcal{D}$, which are necessary at large $b$, are even more difficult to compare.

A common approach is to use the "renormalization-group improved" rapidity anomalous dimension (see e.g.~\cite{Collins:2011zzd,Chiu:2012ir}). In sec.~\ref{sec:improvedD} we demonstrate that such a method corresponds to an evolution along a specifically selected path. It is not the only method to resolve the solution dependence problem, because the path-dependence is caused not by  large logarithms but by the run of coupling constant as it is demonstrated in sec.~\ref{sec:ADinTPT}. The presence of logarithms only amplify the numerical evidence. Therefore, there are two principal solutions for the problem, either to use the commonly defined classes of evolution paths, either to use a  solution that is explicitly independent on the path. We present examples of both methods in sec.~\ref{sec:improvedD} and sec.~\ref{sec:improvedG} respectively.

To our best knowledge such a problem is unique for a double-scale evolution. Clearly, it must be taken into account in phenomenological applications and during the comparison of models and fits. We emphasize that the naive application of resummed rapidity anomalous dimensions does not solve the problem of path dependence of the solution, although it reduces its numerical importance. To control the effects of resummation and guarantee the perturbative convergence for the evolution factor one should take into account the two dimensional nature of TMD evolution. This section is devoted to a detailed description of the effects of truncation of the perturbative series in  TMD evolution, and to disclose the sources of solution dependence.

\subsection{TMD anomalous dimensions in truncated perturbation theory}
\label{sec:ADinTPT}

 We recall that the perturbative expansions for the ultraviolet anomalous dimensions read
\begin{eqnarray}
\Gamma(\mu)=\sum_{n=0}^\infty a_s^{n+1}(\mu)\Gamma_n,\qquad \gamma_V(\mu)=\sum_{n=1}^\infty a_s^n(\mu)\gamma_n,
\end{eqnarray}
where $a_s=g^2/(4\pi)^2$. The leading coefficients in these expansions are $\Gamma_0=4C_F$ and $\gamma_1=-6C_F$ for the quark. In the gluon case, they are $\Gamma_0=4C_A$ and $\gamma_1=-2\beta_0$ (where $\beta_0$ is defined after eq.~(\ref{def:beta})). For the collection of higher order terms see e.g. appendix D in ref.\cite{Echevarria:2016scs}. The perturbative series for the rapidity anomalous dimension $\mathcal{D}$ is
\begin{eqnarray}\label{th:dnk}
\mathcal{D}(\mu,b)=\sum_{n=1}^\infty a_s^n(\mu)\sum_{k=0}^n \mathbf{L}_\mu^k d^{(n,k)},
\end{eqnarray}
where $\mathbf{L}_\mu$ is defined in eq.~(\ref{def:Lmu}) and $d^{(n,k)}$ are numbers. Note, that using  eq.~(\ref{th:dDD=G}) the coefficients $d^{(n,k)}$ with $k>0$ are expressed in the terms of $d^{(i,0)}$, $\Gamma_i$ and the coefficients of $\beta$-function. The leading terms of $\mathcal{D}$ are $d^{(1,1)}=\Gamma_0/2$ and $d^{(1,0)}=0$. The explicit expressions for $d^{(n,k)}$ up to $n=3$ can be found in \cite{Vladimirov:2017ksc}. The running of the coupling constant is given by
\begin{eqnarray}\label{def:beta}
\mu^2 \frac{d a_s(\mu)}{d\mu^2}=-\beta(a_s),\qquad \beta(a_s)=\sum_{n=0}^\infty a_s^{n+2}(\mu)\beta_n,
\end{eqnarray}
where $\beta_0=\frac{11}{3}C_A-\frac{2}{3}N_f$.

In order to study the effects of the truncation of perturbation theory  one has to carefully examine some formally exact relations. In our case the path dependence of the TMD evolution is introduced by the violation of (\ref{th:dDD=G}). Since the relation among anomalous dimensions is spoiled, in the following, we consider $\gamma_V$, $\mathcal{D}$ and $\Gamma$ as three independent functions.

Let us introduce a new  function which accumulates the violation effect, namely
\begin{eqnarray}\label{def:deltaG}
\delta \Gamma(\mu,b)=\Gamma(\mu)-\mu\frac{d \mathcal{D}(\mu,b)}{d\mu}.
\end{eqnarray}
By $\delta \Gamma^{(N)}$ we denote the function $\delta \Gamma$ when the expression for the $\mathcal{D}$ and $\Gamma$ are truncated at $a_s^N$ (inclusive). One can show that
\begin{eqnarray}\label{def:deltaG_N}
\delta \Gamma^{(N)}=2\sum_{n=1}^N\sum_{k=0}^n n \bar \beta_{n-1}(a_s)a_s^{n-1} d^{(n,k)}\mathbf{L}_\mu^k,
\end{eqnarray}
where $\bar \beta_n$ is the $\beta$-function with first $n$ terms removed
\begin{eqnarray}
\bar \beta_n(a_s)=\beta(a_s)-\sum_{k=0}^{n-1}\beta_k a_s^{k+2}.
\end{eqnarray}
For instance, we have
\begin{eqnarray}\label{th:dGamma1}
\delta \Gamma^{(1)}&=&\Gamma_0 \beta(a_s)\mathbf{L}_\mu\sim \mathcal{O}(a_s^2\mathbf{L}_\mu),
\\\label{th:dGamma2}
\delta \Gamma^{(2)}&=&\Gamma_0 \bar \beta_1(a_s)\mathbf{L}_\mu+\beta(a_s)a_s\(\Gamma_0 \beta_0 \mathbf{L}_\mu^2+2 \Gamma_1 \mathbf{L}_\mu+4 d^{(2,0)}\)\sim \mathcal{O}(a_s^3\mathbf{L}_\mu^2).
\end{eqnarray}
In these expressions we take care not to expand the $\beta$-function because in applications it can be of  different perturbative order with respect to the rest of anomalous dimensions.

Given a truncation of the perturbative series at order $N$, the function $\delta \Gamma$ is formally of the next perturbative order. Nonetheless, it is easy to see that its main contribution is always enhanced by  powers of logarithms. In fact, we have
\begin{eqnarray}\label{th:orderEstimation1}
\delta \Gamma^{(N)}\sim \mathcal{O}(a_s^{N+1}\mathbf{L}_\mu^N).
\end{eqnarray}
Therefore, at any (finite) perturbative order there is a region of (large-)$b$ where $\delta \Gamma\sim {\cal  O}(1)$. Moreover, since typically at large $b$ the scale $\mu$ approaches some fixed value, the boundary of the region $\delta \Gamma\sim {\cal O}(1)$ approaches some fixed value for $N\to \infty$. In other words, it is \textit{not possible} to keep $\delta \Gamma$ small by increasing the order of perturbative theory $N$. One can always find the region of $b>b_0$ where $\delta \Gamma^{(N)}\sim {\cal O}(1)$ for any $N$. Clearly, such large values of $b$ correspond to the non-perturbative regime of QCD. Nonetheless, even within a defined model for non-perturbative physics, this ambiguity is present and it should be fixed.

A direct consequence of the violation of eq.~(\ref{th:dDD=G}) is the loss of the integrability condition in eq.~(\ref{th:consitency}) and  consequently the solution of eq.~(\ref{th:TMD_R}) is path-dependent. On top of this, the violation of integrability condition turns into the violation of the transitivity condition of eq.~(\ref{th:transitivity}) and the inversion rule of eq.~(\ref{th:inversion}). For example, for the solutions 1 and 2 we have
\begin{eqnarray}
R[b;\{\mu_1,\zeta_1\}\xrightarrow{1} \{\mu_2,\zeta_2\}]=R^{-1}[b;\{\mu_2,\zeta_2\}\xrightarrow{2} \{\mu_1,\zeta_1\}]\neq 
R^{-1}[b;\{\mu_2,\zeta_2\}\xrightarrow{1} \{\mu_1,\zeta_1\}].
\end{eqnarray}
This demonstrates that if a particular evolution solution has been used for  modeling or fitting,  in order to extend it to a broader interval of energies  one should apply an inverted evolution solution. In turn, this can introduce some extra effects due to the violation of transitivity. It is clear that the effect of solution dependence is proportional to the area between different paths. Therefore, the evolution between well separated scales has an additional enhancement. Specially for this reason the problem of ambiguity should be considered with care before any global fit which would connect high-energy Drell-Yan and low-energy SIDIS data.

To conclude this sub-section we recall that
at small-$b$ the discussed problem could be softened by resumming of the contributions $\sim a_s\mathbf{L}_\mu$, which can be done either implicitly by an improved $\mathcal{D}$ method, which is discussed in sec.~\ref{sec:improvedD}, either explicitly as in~\cite{Echevarria:2012pw} (see also appendix \ref{app:Res}). In this case, we have 
\begin{eqnarray}\label{th:orderEstimation2}
\delta \Gamma^{(N)}\sim \mathcal{O}(a_s^{N+1}\mathbf{L}_\mu).
\end{eqnarray}
Therefore, the integrability condition is still violated but to a smaller extent. Yet the resummation methods are valid only for the regions of $b$ where the non-perturbative effects are negligible. For larger-$b$ some prescription has to be used.

Let us emphasize that \textit{the violation of integrability condition}, and thus the path dependence of evolution, \textit{is not caused by the logarithms} in the rapidity anomalous dimension. The logarithm contributions only amplify the numerical amount of violation and make this effect evident. This argument can be evinced by examining the expression (\ref{th:dGamma2}), which is non-zero even when the logarithmic terms were absent. Therefore, the path dependence problem can not be solved entirely by a resummation of logarithmic contributions. On the contrary, the integrability condition is exactly preserved if the $\beta$-function is zero (i.e. in conformal field theories), even if the value of $\mathbf{L}_\mu$ is large.

\subsection{Formal treatment of TMD evolution in the truncated perturbation theory}

In this section we present the formal treatment of the evolution field in the truncated perturbation theory, where eq.~(\ref{def:irrotational}) does not hold. In other words, the evolution field $\mathbf{E}$ is a \text{non-conservative} vector field. Using the Helmholtz decomposition we split the evolution field into two parts
\begin{eqnarray}
\mathbf{E}(\vec \nu,b)=\tilde{\mathbf{E}}(\vec \nu,b)+\mathbf{\Theta}(\vec \nu,b).
\end{eqnarray}
The fields $\tilde{\mathbf{E}}$ and $\mathbf{\Theta}$ are irrotational and divergence-free respectively,
\begin{eqnarray}
\text{curl}\tilde{\mathbf{E}}=0,\qquad \vec \nabla\cdot \vec \Theta=0,
\end{eqnarray}
where $\text{curl}(\mathbf{curl})=\nabla^2$. They are orthogonal to each other
\begin{eqnarray}
\tilde{\mathbf{E}}\cdot \mathbf{\Theta}=0.
\end{eqnarray}
The irrotational field $\tilde{\mathbf{E}}$ is the conservative part of evolution flow, and can be written as the gradient of a scalar potential
\begin{eqnarray}\label{def:Utilde}
\tilde{\mathbf{E}}(\vec \nu,b)=\vec \nabla \tilde U(\vec \nu,b).
\end{eqnarray}
The divergence-free part in two-dimensions can be written as the vector curl of another scalar potential
\begin{eqnarray}
\mathbf{\Theta}(\vec \nu,b)=\mathbf{curl}\,V(\vec \nu,b),
\end{eqnarray}
where operation $\mathbf{curl}$ is defined in eq.~(\ref{def:grad+curl}).
The curl of the evolution field can be calculated using the definitions 
(\ref{th:dGammaF=G},~\ref{th:dDD=G},~\ref{def:deltaG}),
\begin{eqnarray}
\text{curl}\mathbf{E}=\text{curl}\mathbf{\Theta}=\frac{\delta \Gamma(\vec \nu,b)}{2}\ ,
\end{eqnarray}
and, using  to  Green's theorem, the closed-contour integral of the evolution field is
\begin{eqnarray}
\oint_C \mathbf{E}\cdot d\vec \nu
=\frac{1}{2}\int_\Omega d^2\nu \, \delta \Gamma(\vec \nu,b),
\end{eqnarray}
where $C$ is some closed contour and $\Omega$ is the area surrounded by this contour. Using this expression, we can calculate the difference between solutions   evaluated  on different paths, see eq.~(\ref{th:R_in2D_notation}),
\begin{eqnarray}\label{th:area-relation}
\ln\frac{R[b;\{\mu_1,\zeta_1\}\xrightarrow{P_1} \{\mu_2,\zeta_2\}]}{R[b;\{\mu_1,\zeta_1\}\xrightarrow{P_2} \{\mu_2,\zeta_2\}]}=
\oint_{P_1 \cup P_2} \mathbf{E}\cdot d\vec \nu=\frac{1}{2}\int_{\Omega(P_1 \cup P_2)}d^2 \nu\, \delta \Gamma(\vec \nu,b),
\end{eqnarray}
where ${P_1 \cup P_2}$ is the closed path build from paths $P_1$ and $P_2$ and $\Omega(P_1 \cup P_2)$ is the area surrounded by these paths. 
In turn using the independence of $\delta \Gamma$ on the variable $\zeta$,  eq.~(\ref{def:deltaG}), we can rewrite it as
\begin{eqnarray}
\ln \frac{R[b;\{\mu_1,\zeta_1\}\xrightarrow{P_1} \{\mu_2,\zeta_2\}]}{R[b;\{\mu_1,\zeta_1\}\xrightarrow{P_2} \{\mu_2,\zeta_2\}]}=
\int_{\mu_2}^{\mu_1}\frac{d\mu}{\mu}\delta \Gamma(\mu,b)\ln\(\frac{\zeta_1(\mu)}{\zeta_2(\mu)}\),
\end{eqnarray}
where $\zeta_{1,2}(\mu)$ is the $\zeta$-component of the path $P_{1,2}$ at the scale $\mu$. In the case of solutions 1 and 2, paths are straight and thus $\zeta_{1,2}$ are independent on $\mu$. 
Therefore, comparing solution 1 and 2 we obtain
\begin{eqnarray}\label{ex:sol1-sol2}
\frac{\text{solution 1}}{\text{solution 2}}=\exp\[\ln\(\frac{\zeta_f}{\zeta_i}\)\int_{\mu_i}^{\mu_f}\frac{d\mu}{\mu}\delta\Gamma(\mu,b)\].
\end{eqnarray}
One can see that this expression is enhanced by an extra logarithm of scale separation. This logarithm is typically large, namely $\sim {\cal O}(\mathbf{L}_Q)$. Using the order estimation eq.~(\ref{th:orderEstimation1}) we have
\begin{eqnarray}
\ln\frac{\text{solution 1}}{\text{solution 2}}\sim \mathcal{O}(a_s^N\mathbf{L}^{N+1}_Q).
\end{eqnarray}
In  fig.~\ref{fig:Comapre_lnR} one can observe the difference in the numerical value of eq.~(\ref{ex:sol1-sol2}) comparing red and blue lines. In the resummed case eq.~(\ref{th:orderEstimation2}) one obtains
\begin{eqnarray}
\ln\frac{\text{solution 1}}{\text{solution 2}}\sim \mathcal{O}(a_s^N\mathbf{L}^2_Q).
\end{eqnarray}
These estimations describe the observation that the effect of solution path-dependence is significant, even in the resummed case. Indeed, assuming counting $a_s\mathbf{L}\sim 1$, the difference between solution 1 and 2 is $\sim a_s^{N-2}$ in the resummed case (in the fixed order case it is fixed $\sim a_s^{-1}$). So at $N=3$ (that is indicated as NNLO) the difference is as large as improvement between LO and NLO, which is clearly seen in fig.~\ref{fig:Comapre_lnR}.

\section{Restoration of path-independence}
\label{sec:restoreP}

From the discussion above  one can infer that the path-independence  of the TMD evolution passes through the conservation of the evolution flow field $\mathbf{E}$. 

One possibility to achieve it consists in modifying the evolution field  such that the divergence-free component vanishes and, as a result, only the curl-free component enters in the evolution factor. The expression for the TMD evolution factor has the potential form (compare to eq.~(\ref{th:potential-solution}))
\begin{eqnarray}
\ln R[b;\vec \nu_f\to\vec \nu_i]=\tilde U(\vec \nu_f,b)-\tilde U(\vec \nu_i,b),
\end{eqnarray}
where $\tilde U$ is the scalar potential determined by $\tilde{\mathbf{E}}$, eq.~(\ref{def:Utilde}). In general, the potential $\tilde U$ does not coincide with the potential $U$ defined in eq.~(\ref{th:potential_explicit}). Moreover,  the scalar potential $U$ satisfies the gradient equation (\ref{def:scalarP}), while, in contrast, the scalar potential $\tilde U$ satisfies the Poisson equation
\begin{eqnarray}\label{th:Poisson}
\nabla^2 \tilde U(\vec \nu,b)=\frac{1}{2}\frac{d \gamma_F(\vec \nu)}{d\nu_1}.
\end{eqnarray}
Consequently, the potential $\tilde U$ can be fixed  only up to an  arbitrary harmonic function $f(\vec \nu)$ (with  $\nabla^2f=0$). To fix this ambiguity, an additional statement on the field $\mathbf{\tilde {E}}$ is required, e.g. a boundary condition on a line. Such a boundary condition is equivalent to  imposing a null value of the  divergence-free component $\mathbf{\Theta}$. Unfortunately, nowadays, any statement on the non-perturbative behavior of $\mathcal{D}$ is mostly a conjecture. 

In this work instead we pursue a different strategy. Instead of defining the boundary condition for the eq.~(\ref{th:Poisson}), we repair the compatibility condition in eq.~(\ref{th:consitency}) by improving the definition of anomalous dimensions $\gamma_F$ and/or $\mathcal{D}$ with  terms  of higher-perturbative order. Of course this improvement is not unique, so that here we explore the cases where only one of these anomalous dimensions  is changed. In the following section we consider both scenarios, and call them improved  $\mathcal{D}$ in sec.~\ref{sec:improvedD} and improved  $\gamma$ scenarios, sec.~\ref{sec:improvedG}. Of course, both these scenarios are equivalent to a particular selection of the scalar potential $\tilde U$.

\subsection{Improved $\mathcal{D}$ scenario}
\label{sec:improvedD}
In order  to fix the features of this scenario one observes that the relation (\ref{th:dDD=G}) can be used as an exact relation, i.e. in order to guarantee it to all orders, we replace the perturbative expression for $\mathcal{D}$ by the solution of (\ref{th:dDD=G}). In this  way one obtains
\begin{eqnarray}\label{def:improvedD}
\mathcal{D}(\mu,b)=\int^{\mu}_{\mu_0}\frac{d\mu'}{\mu'}\Gamma(\mu')+\mathcal{D}(\mu_0,b).
\end{eqnarray}

In the improved $\mathcal{D}$ picture the scalar potential $\tilde U$ is obtained from eq.~(\ref{th:potential_explicit}) replacing $\mathcal{D}$ by eq.~(\ref{def:improvedD}). It reads
\begin{eqnarray}
\tilde U(\vec \nu,b;\mu_0)=\int^{\nu_1}_{\ln \mu^2_0}\frac{\Gamma(s)(s-\nu_2)-\gamma_V(s)}{2}ds-\mathcal{D}(\mu_0,b)\nu_2+\const(b).
\end{eqnarray}
One can demonstrate that this approach is equivalent to imposing to the solution of the Poisson equation eq.~(\ref{th:Poisson}) the condition
\begin{eqnarray}\label{imD:mu0}
\delta \Gamma(\mu_0,b)=0.
\end{eqnarray} The expression for the corresponding TMD evolution factor depends on $\mu_0$ and reads
\begin{eqnarray}\label{th:RimprovedD}
\text{improved $\mathcal{D}$ solution:}\qquad \ln R[b;(\mu_f,\zeta_f)\to(\mu_i,\zeta_i);\mu_0]&=&\int^{\mu_f}_{\mu_i}\frac{d\mu}{\mu}\(\Gamma(\mu)\ln\(\frac{\mu^2}{\zeta_f}\)-\gamma_V(\mu)\)
\\\nn &&-\int_{\mu_0}^{\mu_i}\frac{d\mu}{\mu}\Gamma(\mu)\ln\(\frac{\zeta_f}{\zeta_i}\)-\mathcal{D}(\mu_0,b)\ln\(\frac{\zeta_f}{\zeta_i}\).
\end{eqnarray}
Comparing the improved $\mathcal{D}$ solution with the solutions 1 and 2 in eq.~(\ref{th:Rpath1},~\ref{th:Rpath2}) we conclude that it corresponds to a composition of solution 1  and 2 in the usual implementation of TMD evolution
\begin{eqnarray}
R[b;(\mu_f,\zeta_f)\to(\mu_i,\zeta_i);\mu_0]&=&
R[b;(\mu_f,\zeta_f)\xrightarrow{1}(\mu_0,\zeta_0)]
R[b;(\mu_0,\zeta_0)\xrightarrow{2}(\mu_i,\zeta_i)],
\end{eqnarray}
where $\zeta_0$ is arbitrary. The integration path of the improved $\mathcal{D}$ solution is shown in fig~\ref{fig:solutions} by blue lines. The improved $\mathcal{D}$ solution satisfies transitivity and inversion relation eq.~(\ref{th:transitivity},~\ref{th:inversion}), and at $\mu_0=\mu_i(\mu_f)$ it turns into the solution 1,  eq.~(\ref{th:Rpath1}) (into the solution 2, eq.~(\ref{th:Rpath2})). 

The improved $\mathcal{D}$ scenario, is often used in the literature in different forms. For instance, the equation (\ref{def:improvedD}) is used for the resummation of logarithmic contributions within $\mathcal{D}$ in  \cite{Collins:1981uk,Collins:2011zzd,Aybat:2011zv,Scimemi:2017etj,Echevarria:2012pw,Chiu:2012ir,Li:2016axz}. In these cases one has to select $\mu_0$ such that the effect of logarithms in $\mathcal{D}$ is minimized,  that is, typically $\mu_0\sim b^{-1}$ at small-$b$.

Since the improved $\mathcal{D}$ solution is a composition of solutions 1 and 2, it can be seen as the convention for the fixation of a common path for all evolution procedures which depends  on the choice of $\mu_0$. Once the convention for $\mu_0$ is established the comparison of different fits and models is plain. For instance one  can propose  to accept the solution  of eq.~(\ref{imD:mu0}) as a basic agreement. Nevertheless in the absence of such an accepted convention, the improved $\mathcal{D}$ solution should be considered with caution because the numerical differences between different $\mu_0$ could be large. It can be seen already in fig.~\ref{fig:Comapre_lnR}, where the initial scale is selected as $\mu_i\sim b^{-1}$, and thus fulfills the requirement for logarithm minimization. The solution 1 corresponds to (\ref{def:improvedD}) with $\mu_i=\mu_0$. The blue band on it corresponds to variation of $\mu_0\in[\mu_i/,2\mu_i]$ and all these values have reduced logarithm contributions. The width of the band reduces with the increase of perturbative order, but  it is still non-negligible at the highest available order.

\subsection{Improved $\gamma$ scenario}
\label{sec:improvedG}

The integrability condition in  eq.~(\ref{th:consitency}) can be fixed  modifying the anomalous dimension $\gamma_F$ and without using directly eq.~(\ref{th:dGammaF=G}). In this way, one changes the value of the higher order terms in $\gamma_F$. The modified value of $\gamma_F$ (that in the following is denoted by $\gamma_M$) is dependent on $b$, and reads
\begin{eqnarray}\label{def:improvedGamma}
\gamma_M(\mu,\zeta,b)=(\Gamma(\mu)-\delta \Gamma(\mu,b))\ln\(\frac{\mu^2}{\zeta}\)-\gamma_V(\mu).
\end{eqnarray}
The corresponding scalar potential $\tilde U$ is obtained from eq.~(\ref{th:potential_explicit}) by the replacement of $\Gamma\to \Gamma-\delta\Gamma$,
\begin{eqnarray}
\tilde U(\vec \nu,b)=\int^{\nu_{1}} \frac{(\Gamma(s)-\delta\Gamma(s,b))s-\gamma_V(s)}{2}ds-\mathcal{D}(\vec \nu,b)\nu_2+\const(b).
\end{eqnarray}
Using the definition of $\delta \Gamma$, eq.~(\ref{def:deltaG}) and integrating by parts we rewrite this expression in a notably simpler form
\begin{eqnarray}
\tilde U(\vec \nu,b)=-\int^{\nu_{1}} \(\mathcal{D}(s,b)+\frac{\gamma_V(s)}{2}\)ds+\mathcal{D}(\vec \nu,b)(\nu_1-\nu_2)+\const(b).
\end{eqnarray}
Therefore, the corresponding solution for the evolution  factor reads
\begin{eqnarray}\label{th:RimprovedG}
\text{improved $\gamma$ solution:}\qquad \ln R[b;(\mu_f,\zeta_f)\to(\mu_i,\zeta_i)]&=&-\int^{\mu_f}_{\mu_i}\frac{d\mu}{\mu}\(2\mathcal{D}(\mu,b)+\gamma_V(\mu)\)
\\\nn &&+\mathcal{D}(\mu_f,b)\ln\(\frac{\mu_f^2}{\zeta_f}\)-\mathcal{D}(\mu_i,b)\ln\(\frac{\mu_i^2}{\zeta_i}\).
\end{eqnarray}
The expression in eq.~(\ref{th:RimprovedG}) is exceptionally simple, and it explicitly satisfies the transitivity and inversion relations eq.~(\ref{th:transitivity}, \ref{th:inversion}). We stress that this solution is independent of any intermediate points (like $\mu_0$ in the improved  $\mathcal{D}$ case) so that  one does not have  to rely on a common convention for this intermediate point. This is a clear advantage of the improved $\gamma$ scenario in comparison to the more traditional improved $\mathcal{D}$ scenario. All these advantages are also true when the values of $\mathcal{D}$ is modified (e.g. by a non-perturbative contribution). 

Note that for practical applications one has to take care of the logarithmic contributions within $\mathcal{D}$. In contrast to improved $\mathcal{D}$ scenario, where the logarithmic contributions were effectively resummed by the selection of scale $\mu_0$, the improved $\gamma$ scenario does not include any explicit resummation. Therefore, the rapidity anomalous dimension should be taken in a resummed form, e.g. by means of renormalization group (\ref{def:improvedD}) or by explicit resummation (\ref{app:DasX}).

\subsection{The evolution at large-$b$}
\label{sec:NPD}

For large-$b$ the perturbative expansion of $\mathcal{D}$ is not valid. The range of validity of the perturbative expansion $b<\bar b$ can be determined by different methods. One can use the resummed expression~\cite{Echevarria:2012pw} (see also the appendix \ref{app:Res}) and determine the position of the singularity in it. At the leading order the singularity happens at $X=\beta_0 a_s(\mu)\mathbf{L}_\mu=1$. Within such determination the value of $\bar b$ depends on $\mu$ (see the discussion on the behavior of this value in~\cite{Echevarria:2012pw}) and it does not give a clear indication of the perturbative domain. Another way to fix the range of validity of the perturbative series of the  $\mathcal{D}$  function is to consider the stability of the resummed large-$\beta_0$ series as it was done in~\cite{Scimemi:2016ffw}. The analysis made in ref.~\cite{Scimemi:2016ffw} demonstrates that the boundary of perturbative region is $\bar b \sim 3-4$ GeV$^{-1}$. 

In the present  framework we observe that  there exists another natural definition of $\bar b$, as the value at which $\mu_{\text{saddle}}<\Lambda$. This value is $\bar b \sim 3.5$ GeV$^{-1}$, and thus practically coincides with the renormalon estimation~\cite{Scimemi:2016ffw}.

At large-$b$ the shape of the rapidity anomalous dimension is unknown. In fact, the only known information about non-perturbative structure of $\mathcal{D}$ is that it receives renormalon correction $\sim b^2$~\cite{Scimemi:2016ffw,Korchemsky:1994is} (see also~\cite{Becher:2013iya}). It is clear that this contribution is only the first one of a  series of power corrections. So, at large-$b$ the expression for $\mathcal{D}$  should be extracted from data fitting, while at small-$b$ it should match the perturbative expression. Practically the passage from the perturbative  to the non-perturbative regime can be done, e.g.,  by a simple modification
\begin{eqnarray}\label{def:b*}
\mathcal{D}_{\text{NP}}(\mu,b)=\mathcal{D}(\mu,b^*), \qquad b^*(b)=\left\{\begin{array}{cc} b,& b\ll \bar b, \\ b_{\text{max}}, & b\gg \bar b,
\end{array}\right.
\end{eqnarray}
where $b_{\max}$ is a parameter, such that $b_{\text{max}}<\bar b$. An example of such a form  for the non-perturbative correction for rapidity anomalous dimension has been suggested a long  ago in~\cite{Collins:1981va},
\begin{eqnarray}
b^*(b)=b\(1+\frac{b^2}{b_{\max}^2}\)^{-1/2},
\end{eqnarray}
as part of the $b^*$ prescription~\cite{Collins:2011zzd}. Let us stress that the choice of a $b^*$ can be admissible separately for the evolution factor and that  eq.~(\ref{def:b*}) does not imply $b^*$-prescription for the whole TMD distribution.

With the choice $b_{\text{max}}<\bar b$ the saddle point is always in the observable region, which (as it is discussed in the section \ref{sec:zeta-prescription}) allows to determine the optimal TMD.

We note that at large-$b$ the derivative of $\mathcal{D}_{\text{NP}}$ determines the function $\delta \Gamma_{\text{NP}}$. I.e.
\begin{eqnarray}
\delta \Gamma_{\text{NP}}(\mu,b)=\Gamma(\mu)-\mu\frac{d \mathcal{D}_{\text{NP}}(\mu,b)}{d\mu}.
\end{eqnarray}
In the model in eq.~(\ref{def:b*}) it is equal to $\delta \Gamma(\mu,b^*)$. Note, that $\delta \Gamma_{\text{NP}}$ is smaller at large-$b$ since there is no $\mathbf{L}_\mu$ to blow up. Therefore, given the non-perturbative model the problem of solution path-dependence is weakened, and the improved $\mathcal{D}$ and improved $\gamma$ solutions converge to the same. Practically, the implementation of non-perturbative modification of evolution consists in a replacement of $\mathcal{D}$ in the formulas of previous the sections by $\mathcal{D}_{\text{NP}}$.

\section{$\zeta$-prescription and optimal TMD distribution}
\label{sec:zeta-prescription}

The proper construction for the TMD evolution factor is only an (important) piece of the TMD evolution implementation. Another (important) piece is the selection of initial scales for the TMD distribution model. In this section we demonstrate that this problem has a natural solution, that we call the $\zeta$-prescription. In sec.~\ref{subsec:zeta1} we introduce the concept and main characteristics of $\zeta$-prescription. In sec.~\ref{sec:match} we provide expressions for matching coefficients in $\zeta$-prescription. Finally in sec.~\ref{sec:UNITMD} we present a particular implementation for $\zeta$-prescription that has some exceptional properties. We call the TMD distribution defined in this particular prescription, the optimal TMD distribution. It is one of main proposals of this work.

\subsection{$\zeta$ prescription}
\label{subsec:zeta1}

The final point of the rapidity evolution, $\zeta_f$ in eq.~(\ref{th:muzetaf}), is as usual dictated by the hard subprocess. On the contrary, the initial value of the rapidity scale $\zeta_i$ should be selected depending on the input for the  non-perturbative behavior of the TMD distribution. In practice the majority of phenomenological models at small values of $b$ match the TMD distribution to the corresponding collinear distribution. This matching guarantees the agreement of model with high-energy data, and determines significant part of the TMD distribution. The expression for small-$b$ matching has the form
\begin{eqnarray}\label{th:smallB}
F_{f\ot k}(x,b;\mu_i,\zeta_i)=\sum_{n}\sum_{f'}C^{(n)}_{f\ot f'}(x,\mathbf{L}_{\mu_i},\mathbf{L}_{\sqrt{\zeta_i}})\otimes f^{(n)}_{f'\ot h}(x,\mu_i),
\end{eqnarray}
where $f$ is PDF or FF, and $C$ is the Wilson coefficient function. For the unpolarized TMDPDF and TMDFFs the coefficient functions are known at NNLO~\cite{Gehrmann:2014yya,Echevarria:2015usa,Echevarria:2016scs}, while for the polarized cases they are know only for twist-2 matching at NLO~\cite{Gutierrez-Reyes:2017glx}. The coefficient function includes the dependence on $b$ within the logarithms $\mathbf{L}_\mu$ and $\mathbf{L}_{\sqrt{\zeta}}$. In this way, the initial scales $(\mu_i,\zeta_i)$ explicitly enter in the TMD modeling.

The traditional choice of initial values used in many studies suggests $\zeta_i=\mu_i^2$, see e.g.~\cite{Collins:2011zzd,Aybat:2011zv,Bacchetta:2017gcc}. While this choice looks natural, it has  some serious drawback which undermines its stability. In particular, this scale choice leaves uncompensated the logarithms $\mathbf{L}_\mu$ in the coefficient function. The remaining logarithms $\mathbf{L}_\mu$ unrestrictedly grow at larger $b$. In this regime the matching in eq.~(\ref{th:smallB}) is not valid, and thus should to be modified. In turn, any non-perturbative modification requires another matching procedure of the large-$b$ non-perturbative regime with the small-$b$ perturbative expansion. An example of such a procedure is offered by \cite{Collins:2011zzd}, where $b^*$-prescription is used as a non-perturbative modification of eq.~(\ref{th:smallB}). We remark that such a procedure has a poor stability in the perturbative-to-non-perturbative transition , due to the fact, that any deviation from the matching scale uncovers the uncompensated logarithms. As a result the scale variation around $\zeta_i=\mu_i^2$, induces some large error-bands. All-in-all, we come to conclusion that the popular choice of initial scales $\zeta_i=\mu_i^2$ is accidental and does not grant any improvement in the understanding of TMD distributions.

The main idea of the $\zeta$-prescription is to use the two-dimensional nature of TMD evolution to improve and to extend the perturbative stability of the small-$b$ expansion to the full range of $b$. This idea has been used in~\cite{Scimemi:2017etj}, where it has been shown that a particular choice of $\zeta_i$ as a function of $\mu_i$ completely eliminates the double logarithms from the coefficient functions. In~\cite{Scimemi:2017etj} the largest known set of Drell-Yan data has been fitted within $\zeta$-prescription, and without any extra non-perturbative matching, which shows the practical success of $\zeta$-prescription.

The $\zeta$-prescription consists in a special choice of $\zeta_i$ value as a function of $\mu$ and $b$. The value of $\zeta_i$ is selected such that the initial-scale TMD distribution is independent on $\mu_i$. We denote the corresponding value of $\zeta_i$ as $\zeta_{\mu_i}(b)$. The function of $\zeta_\mu(b)$ draws a curve on the evolution plane. By definition of $\zeta$-prescription, the TMD distribution does not evolve along this curve, and thus it is one of the null-evolution curves defined in sec.~\ref{sec:null-evolution}. Therefore,  the expression for a TMD distribution in the $\zeta$-prescription reads
\begin{eqnarray}\label{th:Finzeta}
F(x,b;\mu_f,\zeta_f)=R[b;(\mu_f,\zeta_f)\to (\mu_i,\zeta_{\mu_i}(\vec \nu_B,b))]F(x,b;\vec \nu_B),
\end{eqnarray}
where $\zeta_\mu$ is defined such that $(\mu_i,\zeta_{\mu_i}(\vec \nu_B,b))\in \omega(\vec \nu_B,b)$. 

Note, that the point $\vec \nu_B$ in eq.~(\ref{th:Finzeta}) just represents a label. It only indicates the selected null-evolution curve, but does not enter the function $F(x,b;\vec \nu_B)$ explicitly. In other words, in  eq.~(\ref{th:Finzeta}) the scale $\vec \nu_B$ can be changed to another scale $\vec\nu_B'$, as long as $\vec\nu_B'$ belong to the same null-evolution curve,
\begin{eqnarray}
F(x,b;\vec \nu_B)=F(x,b;\vec \nu_B'),\qquad \vec \nu_B'\in \vec\omega(\vec \nu_B,b).
\end{eqnarray}
In this sense, {\bf  instead of labeling a TMD distribution by a two parameter label $(\mu_i,\zeta_i)$, we can specify a single parameter label, given by an equipotential curve $\vec \nu_B$}. To emphasize this concept we use the single argument $\vec \nu_B$ in the notation of TMD distribution, eq.~(\ref{th:Finzeta}).

Since the single-labelled TMD distributions depend only on the selected null-evolution curve the value of the initial scale $\mu_i$ is irrelevant (as far as it belongs to a selected null-evolution curve). In particular, it allows to eliminate the parameter $\mu_i$ from  error analysis,
\begin{eqnarray}\label{th:Finzeta_fixedmu}
F(x,b;\mu_f,\zeta_f)=R[b;(\mu_f,\zeta_f)\to (\mu_f,\zeta_{\mu_f}(\vec \nu_B,b))]F(x,b;\vec \nu_B),
\end{eqnarray}
which is equivalent to the evolution along the path of the fixed-$\mu$ solution in eq.~(\ref{th:fixedMu-path}). Obviously, such a form is very convenient because the scale $\mu_f$ is related to the hard scale $Q$, and thus the evolution exponent is entirely perturbative. Additionally, the explicit form of the fixed-$\mu$ solution is notably simpler, see eq.~(\ref{th:R-fixedmu}).

The passage from one null-evolution line to another can be done using the TMD  evolution. We have
\begin{eqnarray}\label{th:fromonecurvetoanother}
F(x,b;\vec \nu_B)=R[b;\vec \nu_B\to \vec \nu_B']F(x,b;\vec\nu_B').
\end{eqnarray}
Here the TMD evolution factor is a universal constant that measures the difference between potentials of null-evolution curves. In  sec.~\ref{sec:UNITMD} we show that when performing a  TMD modeling, there is a preferred choice for the null-evolution curve  namely $\vec \nu_B=\vec \nu_{\text{saddle}}$.  This choice defines the  optimal TMD distribution. 

{\bf The $\zeta$-prescription separates the modeling of the TMD distribution from the factorization procedure.} This is the central feature of $\zeta$-prescription, which is absent in formulations of TMD factorization used before. In non-$\zeta$-prescription formulation the TMD distribution has a $\mu$-dependence that is typically related to the scale $b$. Thus the evolution, and hence non-perturbative modification of $\mathcal{D}$, is somehow incorporated into the model for the TMD distribution. This fact makes difficult and sometimes impossible the comparison among different TMD non-perturbative estimations such as lattice or low-energy effective theories.

The $\zeta$-prescription is self-consistent only when the evolution field is conservative. If this is not the  case, the $\zeta$-prescription  in principle could not be implemented because equipotential curves could not be defined for non-conservative fields. Therefore, in the truncated perturbation theory (which is the only practically possible case) the improved scenarios should be used. The naive version of $\zeta$-prescription used in ~\cite{Scimemi:2017etj} uses the improved $\mathcal{D}$ scenario with $\mu_0=\mu_i$, and thus it is not entirely consistent. Additionally, the naive $\zeta$-prescription in~\cite{Scimemi:2017etj} also uses the perturbative series for the definition of the null-evolution curve, instead of eq.~(\ref{th:equipotential_lines2}), which gives additional inconsistency. These inconsistencies have been somewhat tested by variation of scales $c_{1}$ and $c_3$ (see discussion in sec.~\ref{subsec:tradition}). The corresponding variations give the dominant contribution to~\cite{Scimemi:2017etj} error-band. An updated version of the \texttt{arTeMiDe}~\cite{web} code which removes these inconsistencies  and implements the optimal TMD distributions will be soon released.

\subsection{Matching coefficient in $\zeta$-prescription}
\label{sec:match}

A typical model for TMD distribution incorporates the small-$b$ matching to the collinear functions. The $\zeta$-prescription guarantees that the matching coefficient is free from the double logarithmic contribution, which makes it more stable at larger $b$. In this section we derive the details for the small-$b$ matching coefficient within $\zeta$-prescription. We do not restrict the discussion to some particular quantum numbers of the TMD distributions and collinear distributions, since the general structure is universal. 

The small-$b$ matching has the form of eq.~(\ref{th:smallB}). The label $n$ enumerates the collinear distributions contributing to  small-$b$ OPE at the desired order, which in general, are not restricted to leading twist distributions. The evolution of the collinear distribution $f$ is given by 
\begin{eqnarray}\label{DGLAP}
\mu^2 \frac{d}{d\mu^2}f_{f\ot h}(x,\mu)=\sum_{f'}P_{f\ot f'}(x,\mu)\otimes f_{f'\ot h}(x,\mu),
\end{eqnarray}
where the function $P$ is the splitting function and $f'$ enumerates all intermediate flavors that mix in the matching. For the twist-2 distributions the eq.~(\ref{DGLAP}) is known as DGLAP equation, and the sign $\otimes$ represents the Mellin convolution. 
The distributions of  twist  higher then 2 generally depend on several variables $x_i$. 
In this case, the variable $x$ in eq.~(\ref{DGLAP}) represents a collection of variables and $\otimes$ is an integral convolution over these variables.
 Using  eq.~(\ref{DGLAP}) and the TMD evolution eq.~(\ref{def:TMD_ev_UV},~\ref{def:TMD_ev_RAP}) we derive 
\begin{eqnarray}\label{th:Cev_1}
\mu^2 \frac{d}{d\mu^2} C_{f\ot k}(x,b;\mu,\zeta)&=&\frac{\gamma_F(\mu,\zeta)}{2}C_{f\ot k}(x,b;\mu,\zeta)-\sum_{f'}C_{f\ot f'}(x,b;\mu,\zeta)\otimes P_{f'\ot k}(x,\mu),
\\\label{th:Cev_2}
\zeta \frac{d}{d\zeta} C_{f\ot k}(x,b;\mu,\zeta)&=&-\mathcal{D}^f(\mu,b) C_{f\ot k}(x,b;\mu,\zeta).
\end{eqnarray}
These equations fix the logarithmic part of the coefficient function order-by-order in perturbation theory. The explicit expression for the logarithmic part up to two-loop order can be found e.g. in  ~\cite{Scimemi:2017etj,Echevarria:2015usa,Echevarria:2016scs}.

The value of $\zeta_\mu(b)$ is defined through eq.~(\ref{th:equipotential_eqn_simple}), which explicitly reads as
\begin{eqnarray}\label{th:zeta_eqn_explicit}
\frac{\gamma_F(\mu,\zeta_\mu(b))}{2\mathcal{D}(\mu,b)}=\frac{\mu^2}{\zeta_\mu(b)}\frac{d\zeta_\mu(b)}{d\mu^2}.
\end{eqnarray}
Evaluating equations eq.~(\ref{th:Cev_1},~\ref{th:Cev_2}) at  $\zeta=\zeta_\mu(b)$ and using eq.~(\ref{th:zeta_eqn_explicit}) for simplifications, we obtain
\begin{eqnarray}
\mu^2 \frac{d}{d\mu^2} \hat C_{f\ot k}(x,b;\mu)&=&-\sum_{f'}\hat C_{f\ot f'}(x,b;\mu)\otimes P_{f'\ot k}(x,\mu),
\end{eqnarray}
where $\hat C(x,b;\mu)=C(x,b;\mu,\zeta_\mu(b))$. Thus the perturbative series for the coefficient function has the form (here up to NNLO)
\begin{eqnarray}\label{th:C_in_zeta}
\hat C_{f\ot k}&=&C^{(0)}_{f\ot k}+a_s\(-\mathbf{L}_\mu P^{(1)}_{f\ot k}+C^{(1)}_{f\ot k}+c_1\delta_{fk}\)+a^2_s\Big\{\(\frac{1}{2}P^{(1)}_{f\ot f'}\otimes P^{(1)}_{f'\ot k}-\frac{\beta_0}{2}P^{(1)}_{f\ot k}\)\mathbf{L}_\mu^2
\\\nn&&+\Big[-P^{(2)}_{f\ot k}-\(C^{(1)}_{f\ot f'}+c_1\delta_{ff'}\)\otimes P^{(1)}_{f'\ot k}+\beta_0 C_{f\ot k}^{(1)}\Big]\mathbf{L}_\mu+C_{f\ot k}^{(2)}+c_1 C_{f\ot k}^{(1)}+c_2\delta_{fk}\Big\}+\mathcal{O}(a_s^3),
\end{eqnarray}
where we omit the arguments of the functions as well as the sign for the summation on  $f'$ for brevity. In eq.~ (\ref{th:C_in_zeta}) the functions $P^{(n)}$ and $C^{(n)}$ admit the expansion
\begin{eqnarray}
P(x,\mu)=\sum_{n=1}^\infty a_s^n(\mu)P^{(n)}(x),\qquad C(x,\mathbf{L}_\mu=0,\mathbf{L}_{\sqrt{\zeta}}=0)=\sum_{n=0}^\infty a_s^n(\mu) C^{(n)}(x).
\end{eqnarray}
The constants $c_i$ do not depend on $x$, but they depend on the boundary choice of the equipotential curve,  $\vec \nu_B$.

The dependence on the parameter $\vec \nu_B$ is entirely concentrated in the constants $c_i$. To fix it eq.~(\ref{th:zeta_eqn_explicit}) has to be solved order by order in  perturbation theory. The solution of eq.~(\ref{th:zeta_eqn_explicit}) up to NNLO is
\begin{eqnarray}\label{th:zetaline_PT}
\zeta^{\text{pert}}_\mu(b)&=&C_0\frac{\mu}{b}e^{-v(\mu,b)}
\\\label{def:v}
v(\mu,b)&=&\frac{\gamma_1}{\Gamma_0}+\frac{r_1(b)}{\mathbf{L}_\mu}
\\\nn &&
+a_s(\mu)\Big[\frac{\beta_0}{12}\mathbf{L}_\mu^2+\frac{\gamma_2+d^{(2,0)}}{\Gamma_0}-\frac{\gamma_1\Gamma_1}{\Gamma_0^2}+\frac{r_1(b)\beta_0}{2}+\frac{r_2(b)}{\mathbf{L}_\mu}-\frac{2r_1(b)d^{(2,0)}}{\Gamma_0\mathbf{L}_\mu^2}\Big]
\\\nn &&+a_s^2(\mu)\Big[
\frac{\beta_0^2}{24}\mathbf{L}_\mu^3+\(\beta_1+\frac{\beta_0\Gamma_1}{\Gamma_0}\)\frac{\mathbf{L}_\mu^2}{12}
+\(-\frac{\beta_0\gamma_1\Gamma_1}{\Gamma_0^2}+(8d^{(2,0)}+3\gamma_2)\frac{\beta_0}{3\Gamma_0}+\frac{5}{6}\beta_0^2r_1(b)\)\frac{\mathbf{L}_\mu}{2}
\\\nn&&
+\frac{\gamma_1\Gamma_1^2}{\Gamma_0^3}-\frac{\Gamma_1(d^{(2,0)}+\gamma_2)+\gamma_1\Gamma_2}{\Gamma_0^2}+\frac{d^{(3,0)}+\gamma_3}{\Gamma_0}
+\frac{\beta_0\Gamma_1 r_1(b)}{2\Gamma_0}+\frac{\beta_1 r_1(b)+3\beta_0 r_2(b)}{2}
\\\nn && +\frac{r_3(b)}{\mathbf{L}_\mu}-\(d^{(3,0)}r_1(b)+d^{(2,0)}r_2(b)-\frac{d^{(2,0)}\Gamma_1r_1(b)}{\Gamma_0}\)\frac{2}{\Gamma_0 \mathbf{L}_\mu^2}
+\frac{4 d^{(2,0)} r_1(b)}{\Gamma_0^2 \mathbf{L}_\mu^3}
\Big]
+\mathcal{O}(a_s^3),
\end{eqnarray}
where the constants $r_i$ are defined by the  boundary condition $\zeta_{\mu_B}(b)=\zeta_B$, with $\vec \nu_B=(\ln\mu_B^2,\ln\zeta_B)$. Constants $c_i$ in the Wilson coefficient function (\ref{th:C_in_zeta}) are related to the constants $r_i$ as
\begin{eqnarray}\label{th:c12_forUNITMD}
c_1&=&\frac{r_1\Gamma_0}{2},
\\
c_2&=&\frac{\gamma_1d^{(2,0)}}{\Gamma_0}+\frac{r_1\Gamma_1+r_2\Gamma_0}{2}+\frac{r_1^2\Gamma_0^2}{8}.
\end{eqnarray}
We remark here that the perturbative expression for the equipotential line eq.~(\ref{th:zetaline_PT},~\ref{def:v}) is universal in the sense that it  depends on the quark or gluon origin of the parton, but not on other quantum numbers. 

Within the $\zeta$-prescription, the convolution $C\otimes f$ is more stable at large-$b$, due to the absence of double logarithms and the peculiar functional form of logarithm coefficients. Even in the extreme asymptotic cases the shape (the $x$-dependence) of the convolution $C\otimes f$ does not blow up, but behave as it is expected from the naive probabilistic interpretation of collinear distributions. For example, in the  case of unpolarized distributions the first $x-$moment of the convolution $C\otimes f$ is constant at all orders of perturbative expansion, due to the charge conservation. This fact has been already tested and confirmed in fits of the unpolarized distribution made in ref.~\cite{Scimemi:2017etj}.

\begin{figure}[t]
\centering
\includegraphics[width=0.32\textwidth]{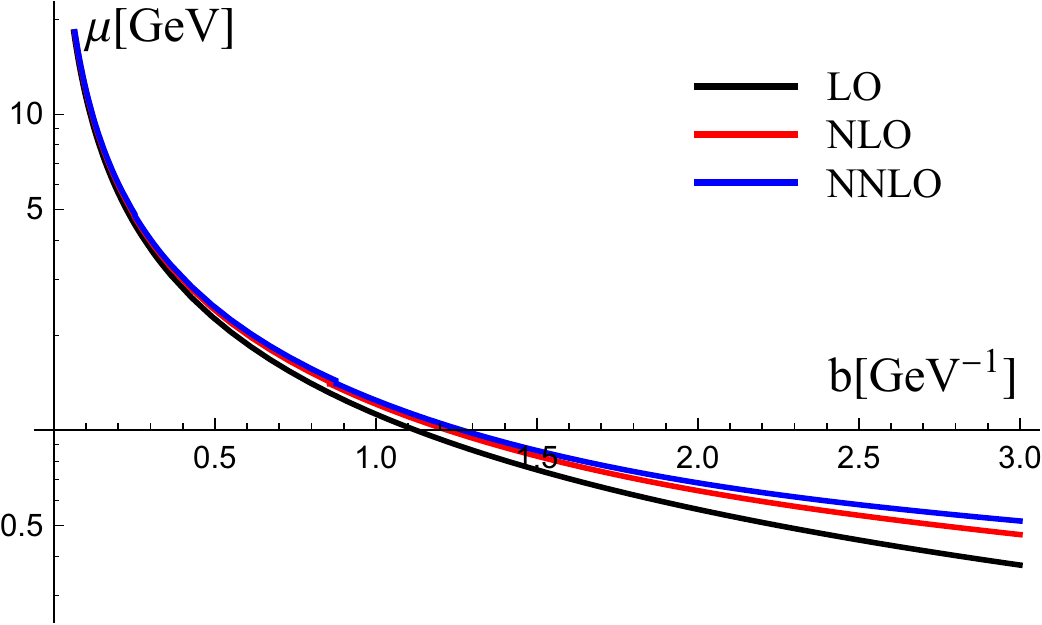}
\includegraphics[width=0.32\textwidth]{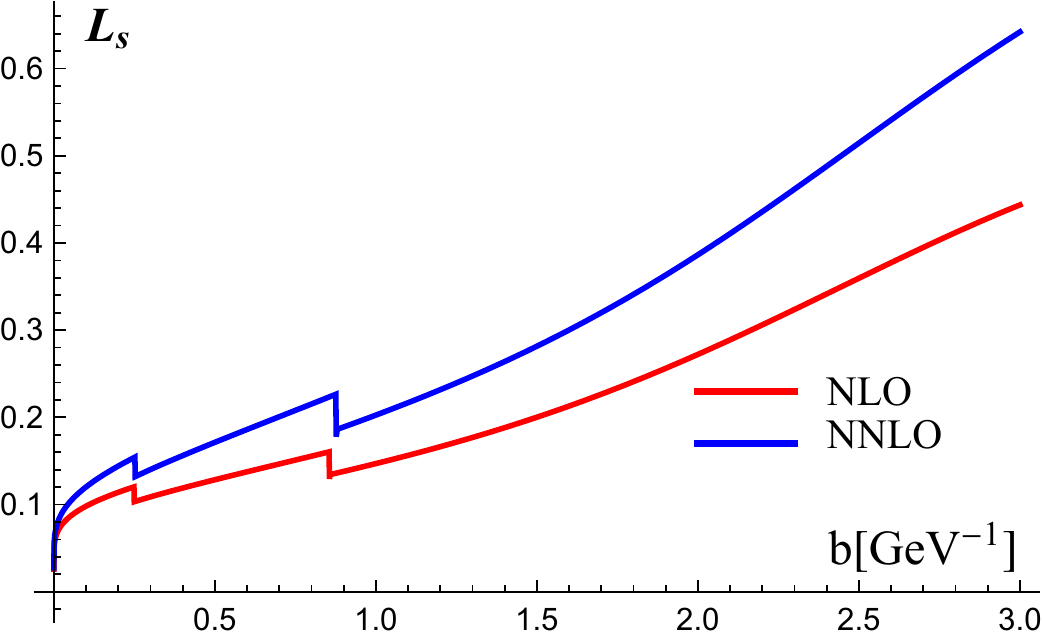}
\caption{(left) Value of $\mu_{\text{saddle}}$ determined by (\ref{def:saddle_values}) at different perturbative orders of $\mathcal{D}$. (right) The value of $\mathbf{L}_s$ at different perturbative orders of $\mathcal{D}$. The LO value of $\mathbf{L}_s$ is exactly zero, and thus is not shown. The kinks are produced by the change on number of active quarks $N_f$ and quark thresholds.}\label{fig:saddle}
\end{figure}

\subsection{Optimal TMD distribution}
\label{sec:UNITMD}

As an outcome of previous section one finds that the coefficient function does not depend on the scale of TMD evolution $\mu_i$.  Instead, the scale that appears  in  the explicit expressions of eq.~(\ref{th:C_in_zeta}), is the intrinsic scale of OPE. To avoid confusion we denote it by $\mu_{\text{OPE}}$. Therefore, the small-$b$ matching of TMD distribution within $\zeta$-prescription has the generic form
\begin{eqnarray}
F_{f\ot k}(x,b;\vec \nu_B)=\sum_{n}\sum_{f'}C^{(n)}_{f\ot f'}(x,b,\vec \nu_B,\mu_{\text{OPE}})\otimes f^{(n)}_{f'\ot h}(x,\mu_{\text{OPE}}).
\end{eqnarray}
The matching scale $\mu_{\text{OPE}}$ is the intrinsic scale of OPE, and is a free parameter. However, its values are restricted to the values of $\mu$ spanned by the defining null-evolution curve. In accordance to the general structure of the evolution plane presented in sec.~\ref{sec:null-evolution}-\ref{sec:singularitiesOfEvPlane}, we have following restrictions on the parameter $\mu_{\text{OPE}}$
\begin{eqnarray}
\text{if }\nu_{B,1}<\ln \mu^2_{\text{saddle}}&\Rightarrow & \mu_{\text{OPE}}<\mu_{\text{saddle}},
\\
\text{if }\nu_{B,1}>\ln \mu^2_{\text{saddle}}&\Rightarrow & \mu_{\text{OPE}}>\mu_{\text{saddle}},
\\
\text{if }\vec \nu_{B}=(\ln \mu^2_{\text{saddle}},\ln\zeta_{\text{saddle}})&\Rightarrow & \mu_{\text{OPE}}\text{ unrestricted}.
\end{eqnarray}
It is clear that the last case is preferable, since the model of TMD distribution is completely unrestricted. Additionally, only this case has a unique definition.

\textbf{The optimal TMD distribution is the distribution defined on this special null-evolution curve.} We denote it simply as $F(x,b)$ emphasizing its scale independence and uniqueness.

The values of $\vec \nu_{\text{saddle}}$ are given by eq.~(\ref{def:saddle_values}). Comparing the second equation eq.~(\ref{def:saddle_values}) with the perturbative expression (\ref{th:zetaline_PT}) we find the values of constants $r_i$
\begin{eqnarray}\label{th:r12}
r_1(b)&=&-\frac{\mathbf{L}_\text{s}^2}{2},\qquad r_2(b)=\frac{\beta_0}{6}\mathbf{L}^3_\text{s}-\frac{2d^{(2,0)}}{\Gamma_0}\mathbf{L}_\text{s},
\\\label{th:r3}
r_3(b)&=&-\mathbf{L}_\text{s}^4\frac{\beta_0^2}{12}+\frac{\mathbf{L}_\text{s}^3}{6}\(\beta_1+\frac{\beta_0\Gamma_1}{\Gamma_0}\)
\\\nn&&+\mathbf{L}_\text{s}^2\frac{\beta_0}{2\Gamma_0}\(4d^{(2,0)}-\gamma_2+\frac{\gamma_1\Gamma_1}{\Gamma_0}\)
-\frac{2\mathbf{L}_\text{s}}{\Gamma_0}\(d^{(3,0)}-\frac{d^{(2,0)}\Gamma_1}{\Gamma_0}\)-\frac{2\{d^{(2,0)}\}^2}{\Gamma_0^2}
\end{eqnarray}
where $\mathbf{L}_\text{s}=\ln(b^2 \mu_{\text{saddle}}^2(b)/4 e^{-2\gamma_E})$. The corresponding values of constants $c_i$ are
\begin{eqnarray}
c_1=-\frac{\Gamma_0}{4}\mathbf{L}_\text{s}^2,
\qquad
c_2=\frac{\Gamma_0^2}{32}\mathbf{L}_\text{s}^4+\frac{\Gamma_0\beta_0}{12}\mathbf{L}_\text{s}^3-\frac{\Gamma_1}{4}\mathbf{L}_\text{s}^2-d^{(2,0)}\mathbf{L}_\text{s}+\frac{d^{(2,0)}\gamma_1}{\Gamma_0}.
\end{eqnarray}
The values of $\mathbf{L}_\text{s}$ could be found by solving the transcendental equation $\mathcal{D}(\mu_\text{saddle},b)=0$. At one loop the solution is $\mathbf{L}_\text{s}=0$ and given in (\ref{th:saddle_crude}). At higher loops this equation can be solved only numerically. The value $\mathbf{L}_s$ slowly grows at larger $b$, but it remains numerically small in comparison to other ingredients of TMD evolution, see fig.~\ref{fig:saddle}. 

Practically, it is inconvenient to have functions $\mathbf{L}_s$ in the coefficient function, because it requires to update the expression for the coefficient function with each correction to the evolution. The more convenient way is to determine the coefficient function on the curve with $r_{1,2}=0$ (which is equivalent to $\mathbf{L}_\text{s}=0$), and take into account the deviation from the exact special null-evolution line by the factor in eq.~(\ref{th:fromonecurvetoanother}). Then the expression for the coefficient function reads
\begin{eqnarray}\label{th:Cpert}
C_{f\ot f'}(x,b;\mu_{\text{OPE}})=\exp\(-\mathcal{D}(\mu,b)\ln\(\frac{\zeta_{\mu}(b)}{\zeta^{\text{pert}}_{\mu}(b)}\)\)C_{f\ot f'}^{\text{pert}}(x,b;\mu_{\text{OPE}}),
\end{eqnarray}
where $C_{f\ot f'}^{\text{pert}}$ is given by eq.~(\ref{th:C_in_zeta}) evaluated on the particular values of $r$ that determine $\zeta^{\text{pert}}_\mu$. In particular, $r_{1,2}=0$. The parameter $\mu$ in eq.~(\ref{th:Cpert}) is a free parameter.

At large-$b$ the saddle point could escape the observable region, i.e. it could appear that $\mu_{\text{saddle}}<\Lambda$. In this case the determination of universal scale-independent TMD distribution is ambiguous, since there is no way to fix  a special null-evolution line.  Of course all this
should be prevented by an appropriate non-perturbative modification of $\mathcal{D}$, as it is discussed in section \ref{sec:NPD}. 

\section{Perturbative uncertainties in TMD factorization}

\begin{figure}[t]
\centering
\includegraphics[width=0.32\textwidth]{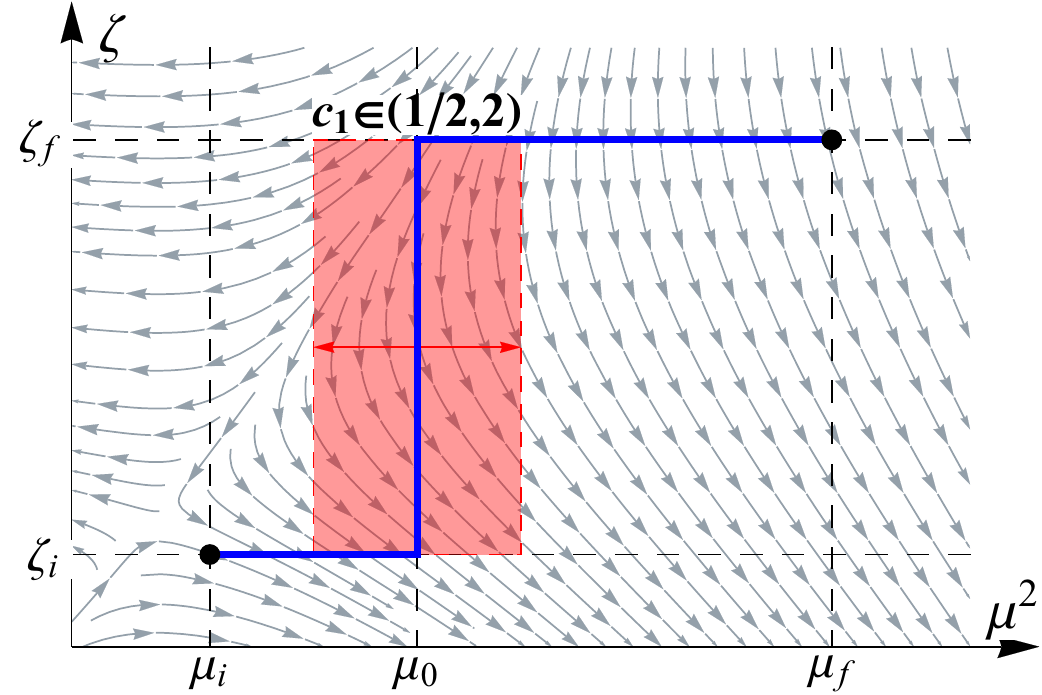}
\includegraphics[width=0.32\textwidth]{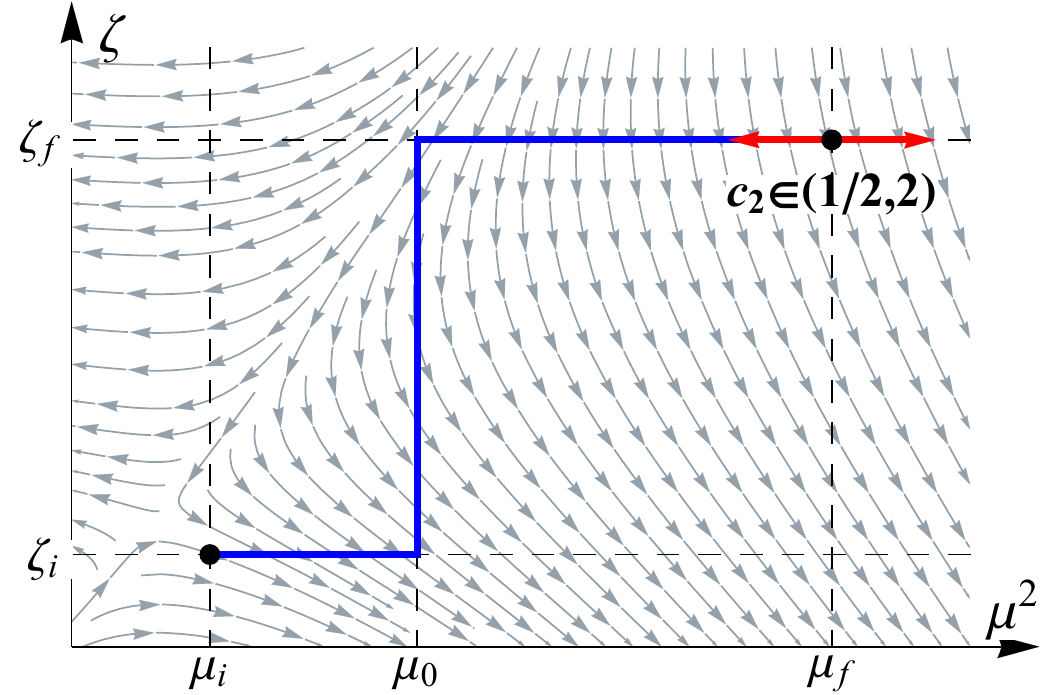}
\includegraphics[width=0.32\textwidth]{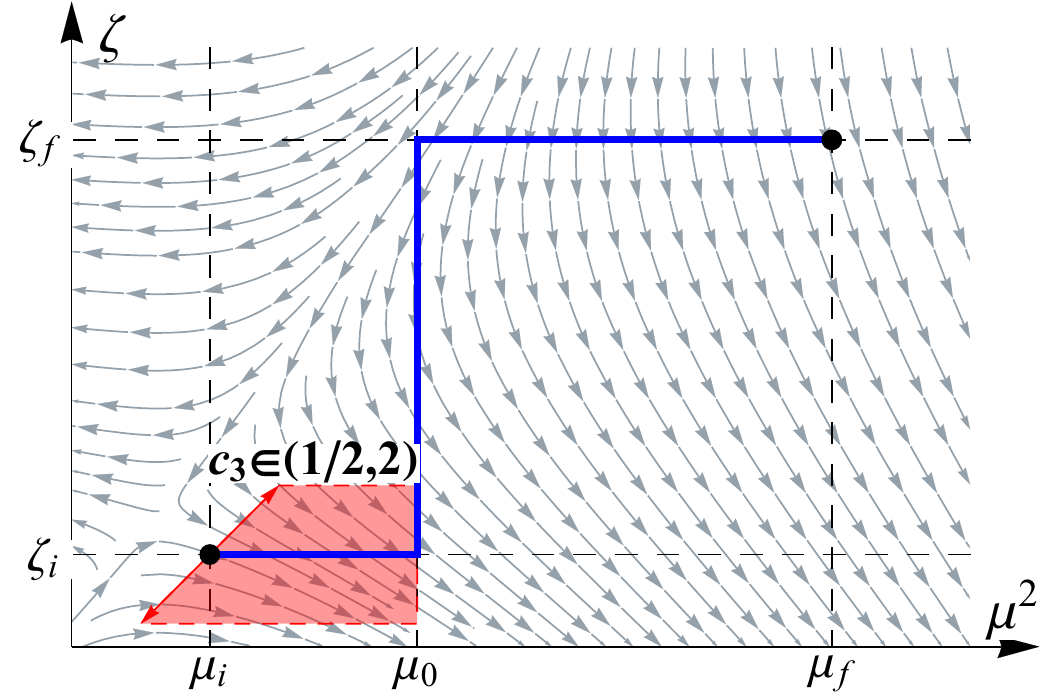}
\caption{Illustration of the path deformation during the variation of parameters $c_1$(left), $c_2$(center) and $c_3$(right). The blue line shows the contour of the evolution in the generic improved $\mathcal{D}$ picture. Red arrows show the displacement of scale positions in the evolution plane during the variation. The red regions show the area which contributes to the solution-dependance (\ref{th:area-relation}). In the present choice of the evolution path, the variation of constant $c_2$ does not deform the solution. The variation of $c_3$ is given for $\zeta_i=\mu_i^2$.} \label{fig:Variation}
\end{figure}

The TMD factorization describes  processes such as Drell-Yan, SIDIS and back-to-back hadron production in $e^+e^-$-annihilation. The factorized expressions for these cross-sections have as common form  the Fourier transformation of a pair of TMD distributions.  In this section we concentrate on the tests of perturbative stability of factorized cross-section that is usually done   varying renormalization/factorization scales. Despite the fact  that the analysis by variations of renormalization scales  have not statistical meaning, it is an important part of the phenomenological studies, since it tests the falsifiability of the theory. Eventual large bands produced by such variations indicate a convergence problem in the perturbative approach and shows the limits of factorization. In this section we demonstrate that the usage of the path-independent solution reduces the variation band, due to absence of the associated path dependence uncertainty. We start this sections recalling in sec.~\ref{subsec:tradition} the  most common inputs for the implementation of TMD inside cross sections. Then in sec.~\ref{subsec:imp-g} we discuss the implementation of the improved $\gamma$-scenario and finally in sec.~\ref{subsec:optimal-s} we  write the cross section using the optimal TMDs. In the following, we show examples with the Drell-Yan cross-section, for definiteness. All the results presented in this section hold for other TMD processes with the appropriate replacements.

\subsection{More traditional implementation of cross-sections within TMD factorization}
\label{subsec:tradition}

Within TMD factorization (and hence at $q_T\ll Q$), the cross-section for Drell-Yan processes has the generic form
\begin{eqnarray}
\frac{d\sigma}{dX}=\sigma_0 \sum_{ff'} \int\frac{d^2 b}{4\pi} e^{i(b\cdot q_T)} H_{ff'}(Q,\mu_f) F_{f\ot h}(x_1,b;\mu_f,\zeta_f)F_{f'\ot h}(x_2,b;\mu_f,\zeta'_f),
\end{eqnarray}
where $dX$ is the $q_T$-differential phase-space element, $\sigma_0$ is the normalization of the cross-section, $H$ is the hard part, and $F$ are TMDPDFs. The values of $x_{1,2}$ are dictated by the kinematics. The parameters $\zeta$ are constrained as $\zeta_f\zeta'_f=Q^4$. It is natural to consider the symmetric point $\zeta_f=\zeta'_f=Q^2$. The scale $\mu_f$ is generically unconstrained but it is selected $\mu_f\sim Q$ in order to minimize the logarithms of $Q/\mu_f$ that are present in the hard coefficient function. Therefore, the final evolution scale of the TMD distributions is $(\mu_f,\zeta_f)=(Q,Q^2)$, as it is discussed in eq.~(\ref{th:muzetaf}).

The model for TMD is made at the initial scale $(\mu_i,\zeta_i)$. The connection between the external-kinematic dependent hard scale and the TMD distribution, is made by the TMD evolution factor. In the improved $\mathcal{D}$ picture eq.~(\ref{th:RimprovedD}), the practical expression for TMD cross-section reads
\begin{eqnarray}\label{xSec:standard}
\frac{d\sigma}{dX}=\sigma_0 \sum_f \int\frac{d^2 b}{4\pi} e^{i(b\cdot q_T)} H_{ff'}(Q,\mu_f) \{R^f[b;(\mu_f,\zeta_f)\to(\mu_i,\zeta_i),\mu_0]\}^2  F_{f\ot h}(x_1,b;\mu_i,\zeta_i)F_{f'\ot h}(x_2,b;\mu_i,\zeta_i),
\end{eqnarray}
where $R$ is defined in eq.~(\ref{th:RimprovedD}). Here, we have used the fact that $R^f=R^{f'}$ as far as both partons are gluons, or quarks. The models for a TMD distribution conventionally include the small-$b$ matching to the integrated distribution supplemented by a non-perturbative function. The typical form is
\begin{eqnarray}
F(x,b;\mu_i,\zeta_i)=\sum_{n}C_n(x,b;\mu_i,\zeta_i;\mu_{\text{OPE}})\otimes f_n(x,\mu_{\text{OPE}}) f_{\text{NP}}(b,x),
\end{eqnarray}
where $C$ is the perturbatively calculable matching coefficient, $f$ is the collinear distribution, and $f_{NP}$ is an ansatz for the non-perturbative large-$b$ behavior  of TMD distribution and it is the object of the fitting procedure. Here, we specially separate the scales of TMD distribution from the scale of OPE, to keep the discussion at the most general level. This is a typical construct used for the phenomenology. The particular details and the choice of scales the implementation vary among authors, compare e.g. realizations used in refs.~\cite{Bacchetta:2017gcc,Scimemi:2017etj,DAlesio:2014mrz,Aybat:2011zv,Bozzi:2010xn,Landry:2002ix}.

In this way the traditional implementation of the TMD cross-section contains four renormalization scale entries of the perturbative series and consequently four scales $\mu$. These are $(\mu_0,\mu_f,\mu_i,\mu_\text{OPE})$. The scales $(\zeta_i,\zeta_f)$ are usually related to $(\mu_i,\mu_f)$ and they are not independent. In the infinitely precise perturbation theory the cross-section \textit{is independent} on each scale $\mu$ \textit{separately} and the residual dependence  on each of  these scales is an artifact of truncation of the perturbative series.

\begin{figure}[t]
\centering
\includegraphics[width=0.95\textwidth]{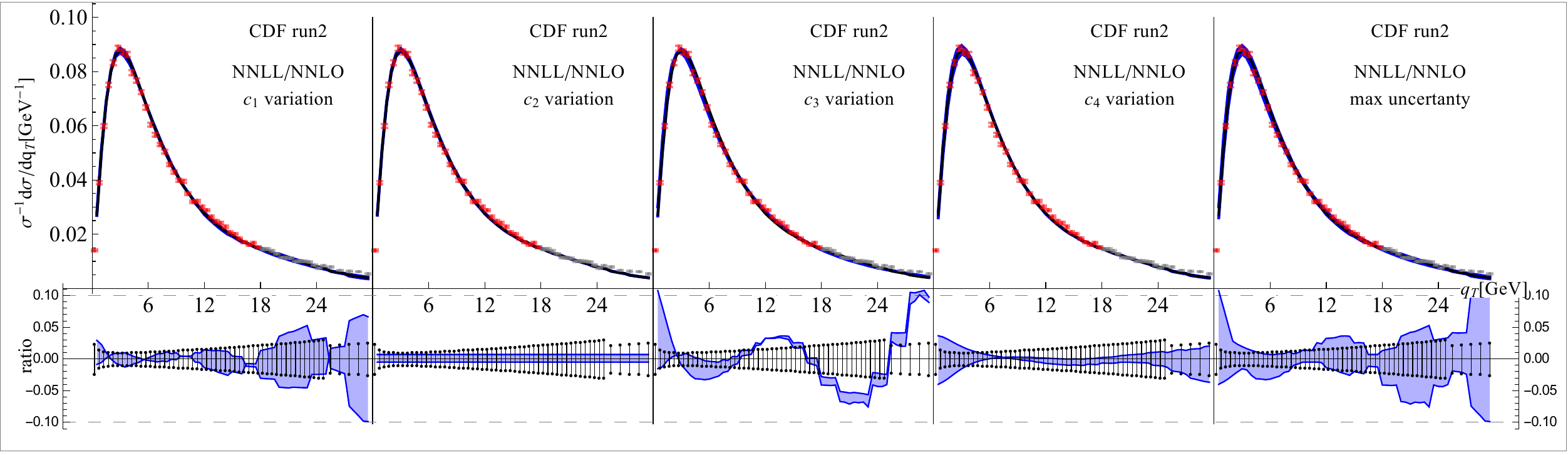}
\caption{Effect of variation of constants $c_{i}$ on the Z-boson production cross-section. The right panel shows the envelope of bands. The picture is from~\cite{Scimemi:2017etj}. For the definition of perturbative orders and other details see~\cite{Scimemi:2017etj}.} \label{fig:ExampleOfVariation}
\end{figure}

The standard method to test the dependence on the scales, and thus the stability of the perturbation theory prediction, is to multiply each scale by a parameter~\cite{Nadolsky:2000ky,Bozzi:2010xn,DAlesio:2014mrz,Scimemi:2017etj} and vary the parameters nearby the central value. E.g. in the notation of~\cite{Scimemi:2017etj}, one changes scales as
\begin{eqnarray}\label{xSec:c1234}
\mu_0\to c_1 \mu_0,\qquad \mu_f\to c_2 \mu_f,\qquad \mu_i\to c_3 \mu_i,\qquad \mu_{\text{OPE}}\to c_4\mu_{\text{OPE}},
\end{eqnarray}
and checks the variations of $c_i\in(1/2,2)$. The variation produces a band which roughly represents the size of the next-perturbative order contribution. Each constant $c_i$ explores a particular theoretical error. The numerical source of the band is the mismatch between resummed (and hence "exact") expression (e.g. TMD evolution factor, or PDF) and the fixed order coefficient function (e.g. hard coefficient function $H$, or small-$b$ matching coefficient). In the TMD cross-section there is an additional source of perturbative scale-dependence, namely, the solution path-dependence. This error is undesirable, since it does not entirely tests the convergence of perturbation theory, and depends on the particular realization of the numerics. Examining the expression (\ref{xSec:standard}) we can sort the variation which test these cases.
\begin{itemize}
\item The variation of $c_1$ tests only the path dependence of evolution. For that reason the variation band for $c_1$ is uniformly large, unstable, and not significantly reducing with order improvement. It is absent in any path-independent solutions.
\item The variation of $c_2$ and $c_3$ tests both the perturbative convergence and the path dependence. The latter is the subject of a particular realization of the evolution exponent. E.g. for the improved $\mathcal{D}$ solution (for well-separated $\mu_0$ and $\mu_f$) the variation of $c_2$ does not deform the path, as it is demonstrated in fig.~\ref{fig:Variation}. Whereas the variation of $c_3$ does deform the path. The effect of it is clearly seen in fig.~\ref{fig:ExampleOfVariation} where the $c_3$ band is dominant (fig.~\ref{fig:ExampleOfVariation} is taken from~\cite{Scimemi:2017etj}, where cross-section has been taken in the form of eq.~(\ref{xSec:standard}).). The usage of a path-independent solution removes this contribution from the $c_2$ and $c_3$ bands leaving only perturbative uncertainty band. 
\item The variation of $c_4$ tests only the perturbative stability. In fact, this scale is not related to TMD factorization and the problems of its implementation.
\end{itemize}
In the following sections, we give explicit expressions for a TMD factorized cross-section in path independent scenarios with and without optimal TMD definition. We also demonstrate, and it is one of the main results of the article, that the variation error-bands improve, in accordance to the general expectations discussed above.

\subsection{The TMD cross-sections with evolution in the improved $\gamma$ picture}
\label{subsec:imp-g}

In order to avoid the undesired ambiguity coming from the solution path-dependence, one can use the improved $\gamma$-picture, suggested in sec.~\ref{sec:improvedG}. In this case the TMD cross-section looks precisely the same as in eq.~(\ref{xSec:standard}), with the only difference that the TMD evolution  factor is taken as in eq.~(\ref{th:RimprovedG}). Let us express these formulas restoring all dropped superscripts for convenience. The TMD cross-section has the form
\begin{eqnarray}\label{xSec:gammaImproved}
\frac{d\sigma}{dX}=\sigma_0 \sum_f \int\frac{d^2 b}{4\pi} e^{i(b\cdot q_T)} H_{ff'}(Q,\mu_f) \{R^f[b;(\mu_f,\zeta_f)\to(\mu_i,\zeta_i)]\}^2  F_{f\ot h}(x_1,b;\mu_i,\zeta_i)F_{f'\ot h}(x_2,b;\mu_i,\zeta_i),
\end{eqnarray}
where
\begin{eqnarray}\label{xSec:RimprovedG}
R^f[b;(\mu_f,\zeta_f)\to(\mu_i,\zeta_i)]&=&\exp\Big\{-\int^{\mu_f}_{\mu_i}\frac{d\mu}{\mu}\(2\mathcal{D}_{\text{NP}}^f(\mu,b)+\gamma^f_V(\mu)\)
\\\nn &&\qquad\qquad+\mathcal{D}_{\text{NP}}^f(\mu_f,b)\ln\(\frac{\mu_f^2}{\zeta_f}\)-\mathcal{D}_{\text{NP}}^f(\mu_i,b)\ln\(\frac{\mu_i^2}{\zeta_i}\)\Big\}.
\end{eqnarray}
The modified rapidity anomalous dimension $\mathcal{D}_{\text{NP}}^f$ is perturbative at small-$b$ and can have non-perturbative correction at large-$b$. In order to improve the perturbative convergence of the anomalous dimension $\mathcal{D}$ the resummed version can also be used, see appendix~\ref{app:Res}.

The cross-section in the improved $\gamma$ picture eq.~(\ref{xSec:gammaImproved},~\ref{xSec:RimprovedG}) is self-consistent in the sense that the incorporated TMD evolution is explicitly transitive, invertible and path independent. 
Therefore, the extractions of TMDs are simple to compare knowing the TMD functions $F$, the non-perturbative evolution $\mathcal{D}_{\text{NP}}^f$ and the scales $(\mu_i,\zeta_i)$ that are used for each extraction. 
The  cancellation of the $\mu$-dependence is achieved adjusting correctly the perturbative orders of ingredients. 
We stress that the typical question about which order of $\Gamma$ one should use in comparison to other anomalous dimensions is absent in this scheme, due to the absence of $\Gamma$. Instead, the rapidity anomalous dimension ${\cal D}$ and $\gamma_V$ should be of the same order, since their finite parts jointly contribute to the integral in eq.~(\ref{xSec:RimprovedG}). In the table~\ref{tab:improvedGamma_orders} we present the consistent order composition in the improved $\gamma$ scheme. 

The test of the perturbative stability can be done in the same manner as usual, i.e. by rescaling the parameters $\mu$ in eq.~(\ref{xSec:c1234}). However now  {\bf  in the improved $\gamma$ picture the parameter $\mu_0$ is absent} by definition. Indeed, the parameter $\mu_0$ and its variation in eq.~(\ref{xSec:standard}) parameterizes and measures the path dependence of the solution (see fig.~\ref{fig:Variation}(left)) only. Therefore, it disappears in the path-independent solution.

\begin{table}[b]
\begin{tabular}{||c|| c | c || c|c||c|c||}
name & ~~$\mathcal{D}$~~ & ~~$\gamma_V$~~ & ~~$H$~~ & ~$C_{f\ot f'}$~ & $a_s$(run) & PDF (evolution) 
\\\hline\hline
LO & $a_s^1$ & $a_s^1$ & $a_s^0$ & $a_s^0$ & lo & lo 
\\\hline
NLO & $a_s^2$ & $a_s^2$ & $a_s^1$ & $a_s^1$ & nlo & nlo 
\\\hline
NNLO & $a_s^3$ & $a_s^3$ & $a_s^2$ & $a_s^2$ & nnlo & nnlo 
\end{tabular}
\caption{\label{tab:improvedGamma_orders} The adjustment of perturbative order for the cross-section in the improved $\gamma$ picture. For explicitness we indicate the highest included power of $a_s$ in the expression.}
\end{table}

\begin{figure}[t]
\centering
\includegraphics[width=0.32\textwidth]{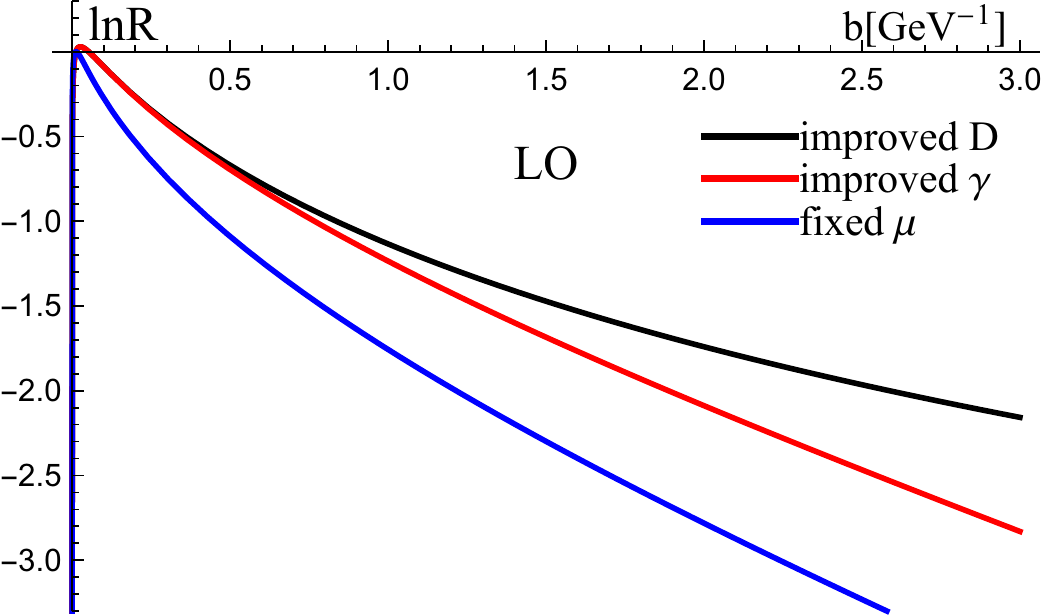}
\includegraphics[width=0.32\textwidth]{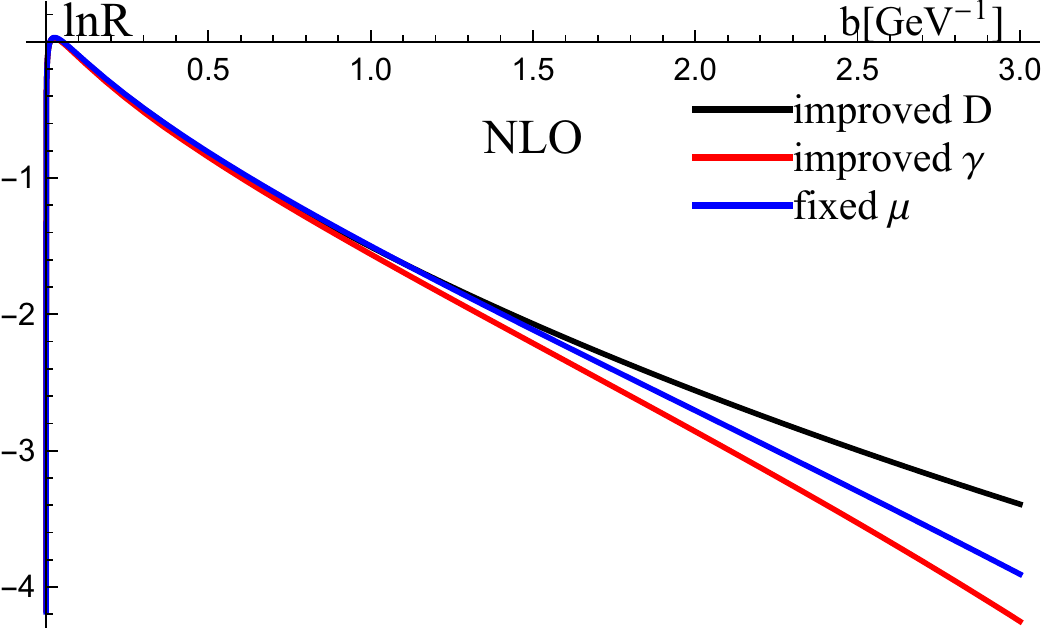}
\includegraphics[width=0.32\textwidth]{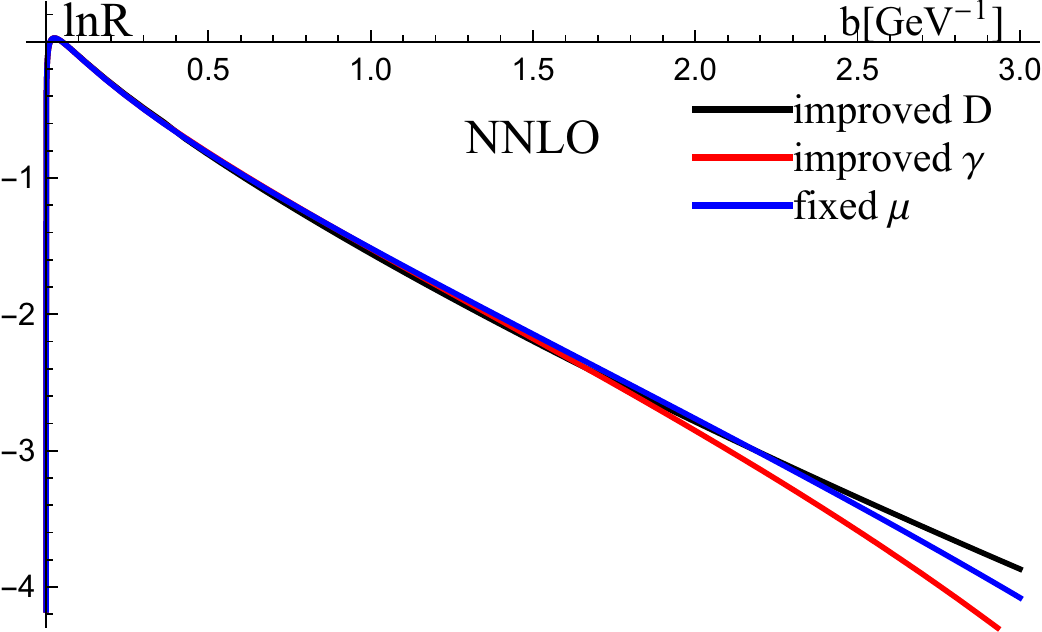}
\caption{Comparison of evolution exponents evaluated in difference schemes. The final point of evolution is $(\mu_f,\zeta_f)=(M_Z,M_Z^2)$. The initial point is $(\mu_i,\zeta_{\mu_i})$ at $\mu_i=C_0/b+2$. The fixed $\mu$-solution indicates the solution (\ref{xSec:UNIR_pert}) with no initial point of evolution. The anomalous dimension $\mathcal{D}$ is taken in the resummed form (\ref{app:DasX}). No non-perturbative modifications of $\mathcal{D}$ are made.} \label{fig:lnRFinal}
\end{figure}

\begin{figure}[t]
\centering
\includegraphics[width=0.32\textwidth]{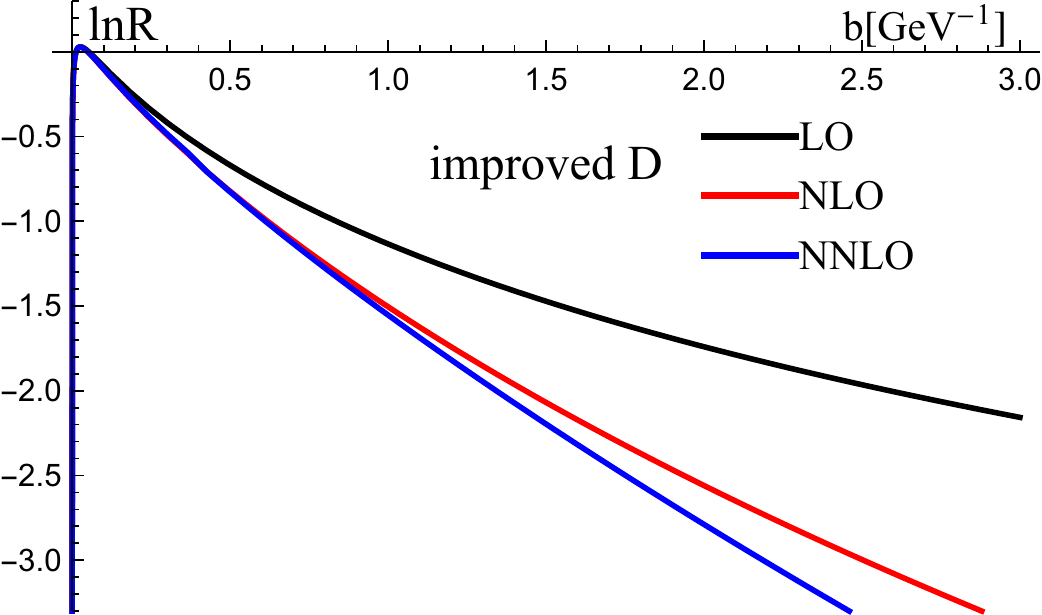}
\includegraphics[width=0.32\textwidth]{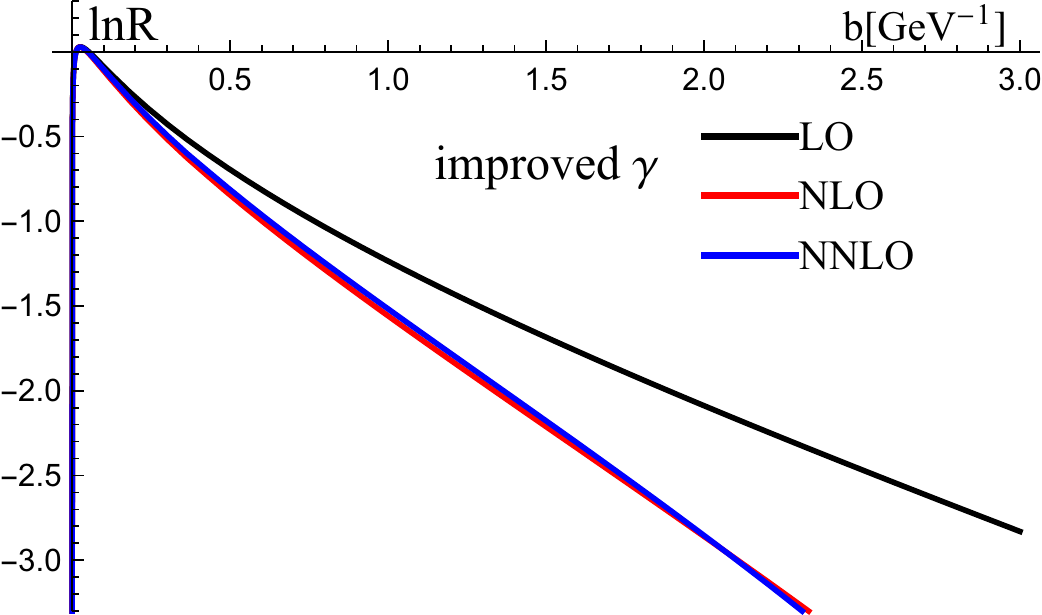}
\includegraphics[width=0.32\textwidth]{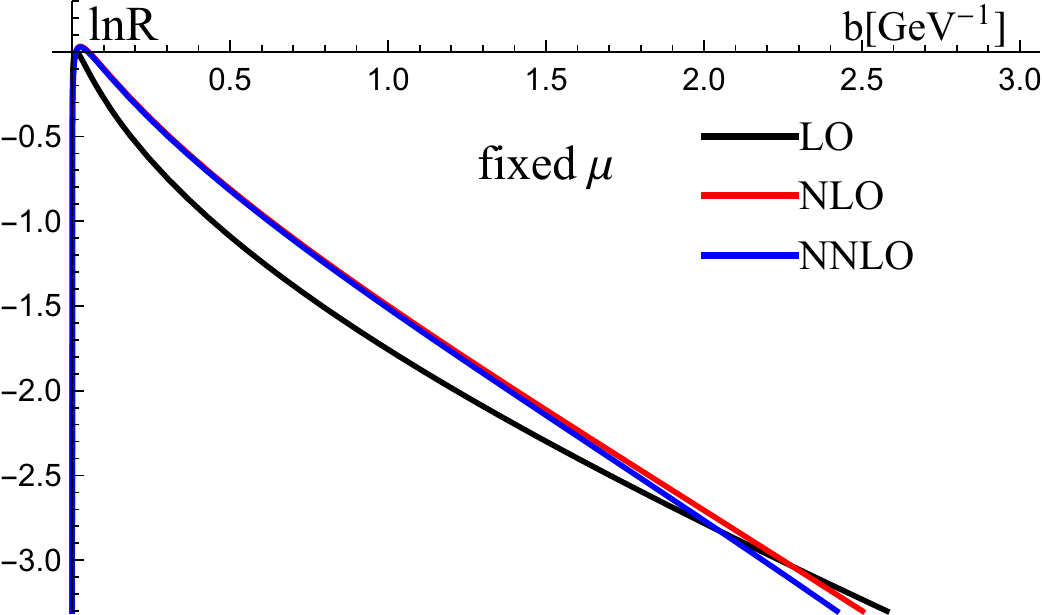}
\caption{Comparison of evolution exponents evaluated at different orders. All inputs are the same as in fig.\ref{fig:lnRFinal}.} \label{fig:lnRFinalEvs}
\end{figure}

\subsection{The TMD cross-sections using the optimal TMD distributions}
\label{subsec:optimal-s}


\begin{figure}[t]
\centering
\includegraphics[width=0.95\textwidth]{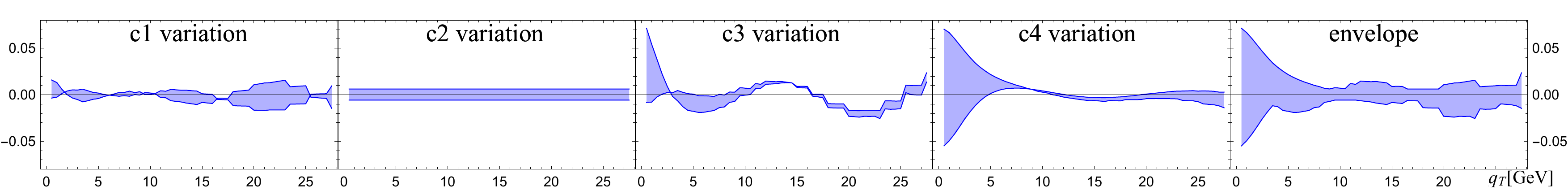}
\includegraphics[width=0.95\textwidth]{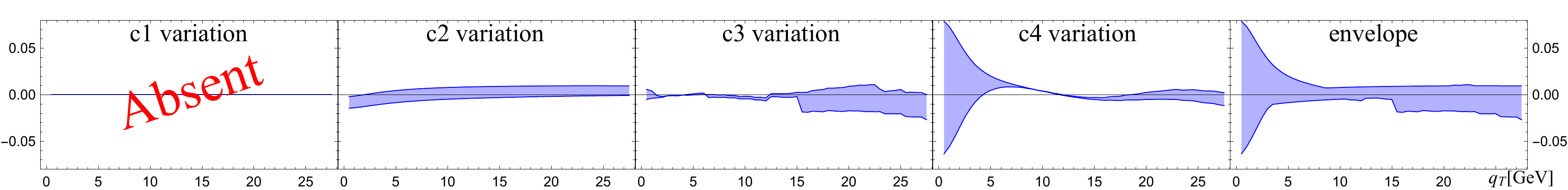}
\includegraphics[width=0.95\textwidth]{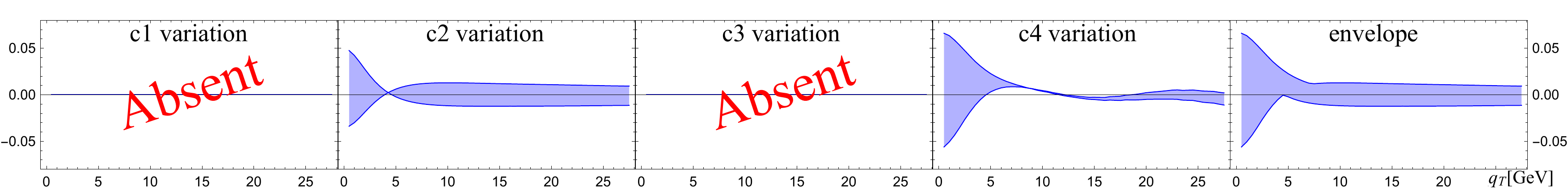}
\caption{Comparison of error bands obtained by the scale-variations for cross-sections given by (\ref{xSec:standard}) (top), (\ref{xSec:gammaImproved}) (middle), (\ref{xSec:UNITMD_pert}) (bottom) at NNLO. Here, the kinematics bin-integration, etc., is for the Z-boson production measure at ATLAS at 8 TeV \cite{Aad:2015auj}.} \label{fig:ATLAS}
\end{figure}

The expression for the cross-section can be simplified even more with the application of the optimal TMD definition discussed in section~\ref{sec:UNITMD}. In this case the TMD cross-section reads
\begin{eqnarray}\label{xSec:UNITMD}
\frac{d\sigma}{dX}=\sigma_0 \sum_f \int\frac{d^2 b}{4\pi} e^{i(b\cdot q_T)} H_{ff'}(Q,\mu_f) \{R^f[b;(\mu_f,\zeta_f)]\}^2  F_{f\ot h}(x_1,b)F_{f'\ot h}(x_2,b),
\end{eqnarray}
where the evolution exponent can be given by two equivalent expressions
\begin{eqnarray}\label{xSec:UNIR}
R^f[b;(\mu_f,\zeta_f)]&=&\exp\Big\{-\int^{\mu_f}_{\mu_{\text{saddle}}}\frac{d\mu}{\mu}\(2\mathcal{D}_{\text{NP}}^f(\mu,b)+\gamma^f_V(\mu)\)
+\mathcal{D}_{\text{NP}}^f(\mu_f,b)\ln\(\frac{\mu_f^2}{\zeta_f}\)\Big\}
\\\label{eq:69}&=&\exp\Big\{-\mathcal{D}_{\text{NP}}^f(\mu_f,b)\ln\(\frac{\zeta_f}{\zeta_{\mu_f}(b)}\)\Big\}.
\end{eqnarray}
where in eq.~(\ref{xSec:UNIR}), the scale $\mu_{\text{saddle}}$ is $b$-dependent, and defined by the equation
\begin{eqnarray}\label{xSec:musaddle}
\mathcal{D}^f_{\text{NP}}(\mu_{\text{saddle}},b)=0.
\end{eqnarray}
The value of $\zeta_\mu(\mu_f,b)$ is defined by eq.~(\ref{th:equipotential_lines2}). 
{\bf The optimal TMD distribution is by definition scale independent}. Its matching coefficient at small-$b$ is given by eq.~(\ref{th:C_in_zeta}) with $c_1$ and $c_2$ defined in eq.~(\ref{th:c12_forUNITMD}). 
We stress that this construction is independent of the type of TMD distribution and of the process, due to the universality of the TMD evolution.

The scheme presented in eq.~(\ref{xSec:UNITMD},~\ref{xSec:UNIR}) is still not very  practical due to the necessity of recalculating  the small-$b$ matching coefficient with each modification of the $\mathcal{D}_{\text{NP}}$. To be more precise, using this implementation is not very costly (in the machine time) at NLO (since coefficient $c_1$ appears only near the $\delta$-function) but it becomes more expensive at higher orders. 

\begin{figure}[t]
\centering
\includegraphics[width=0.95\textwidth]{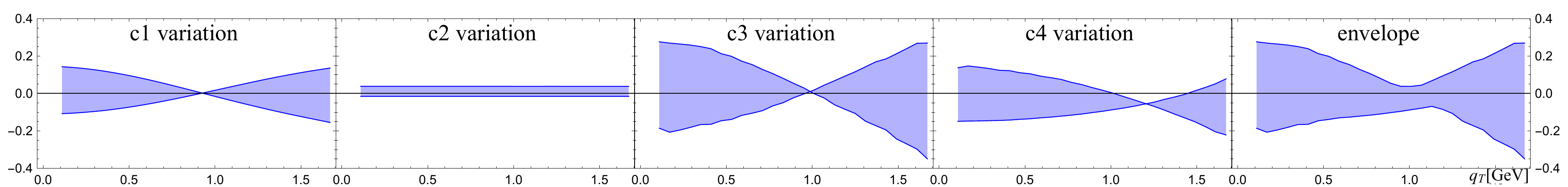}
\includegraphics[width=0.95\textwidth]{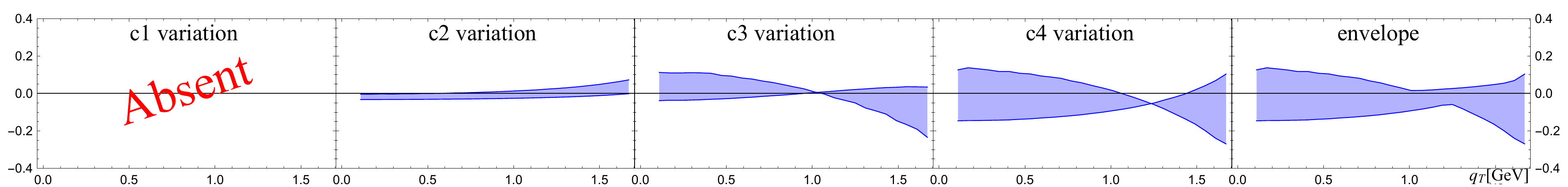}
\includegraphics[width=0.95\textwidth]{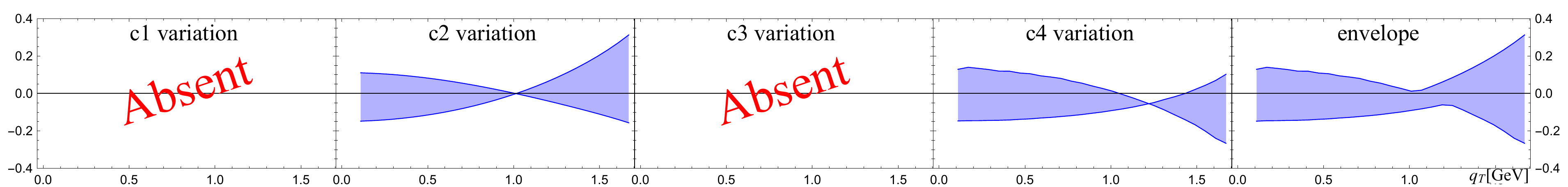}
\caption{Comparison of error bands obtained by the scale-variations for cross-sections given by (\ref{xSec:standard}) (top), (\ref{xSec:gammaImproved}) (middle), (\ref{xSec:UNITMD_pert}) (bottom) at NNLO. Here, the kinematics bin-integration, etc., is for Drell-Yan process measured at E288 experiment at $E_{\text{beam}}=200$GeV and $Q=6-7$GeV \cite{Ito:1980ev}.} \label{fig:E288}
\end{figure}

In order to have  a faster implementation we suggest to exponentiate the boundary constants $r_{i}$ (\ref{th:r12},~\ref{th:r3}), which is equivalent to  switching from the exact special null-evolution line, to the close null-evolution line with $r_i=0$.
In this way we obtain the distribution $\tilde F_{f\ot h}(x,b)$ defined as
\begin{eqnarray}
F_{f\ot h}(x,b)=\exp\Big[-\mathcal{D}_{\text{NP}}^f(\mu,b)\ln\(\frac{\zeta_\mu(b)}{\zeta^{\text{pert}}_\mu(b)}\)\Big]\tilde F_{f\ot h}(x,b)
\end{eqnarray}
Note that, at small-$b$, the condition  $r_i=0$ line coincides with the special line, and at one-loop accuracy they coincide for all values of $b$. The change of the line is to be taken into account by an extra factor in the coefficient function eq.~(\ref{th:Cpert}). This factor is universal and in the fitting expression it can be extracted from the TMD distribution and recombined with the evolution factor $R$. Thus {\bf the practical expression for the optimal TMD cross-section reads}
\begin{eqnarray}\label{xSec:UNITMD_pert}
\frac{d\sigma}{dX}=\sigma_0 \sum_f \int\frac{d^2 b}{4\pi} e^{i(b\cdot q_T)} H_{ff'}(Q,\mu_f) \{\tilde R^f[b;(\mu_f,\zeta_f)]\}^2  \tilde F_{f\ot h}(x_1,b)\tilde F_{f'\ot h}(x_2,b),
\end{eqnarray}
with
\begin{eqnarray}\label{xSec:UNIR_pert}
\tilde R^f[b;(\mu_f,\zeta_f)]&=&\exp\Big\{-\mathcal{D}_{\text{NP}}^f(\mu_f,b)\Big[\ln\(\frac{\zeta_f b}{C_0 \mu_f }\)+v^f(\mu_f,b)\Big]\Big\},
\end{eqnarray}
where $C_0=2e^{-\gamma_E}$ and the function $v$ is given by eq.~(\ref{def:v}) at $r_i=0$. In these expressions we recommend to use the resummed versions of $\zeta^{\text{pert}}_\mu$ and $\mathcal{D}_{\text{NP}}$ at small-$b$ to improve the perturbative convergence. The corresponding expressions are derived in the appendix~\ref{app:Res}. The comparison of $R$ factors  in all three versions of evolution presented here is given in figs.~\ref{fig:lnRFinal}. One can see that at three-loop order the difference among these functions is negligible. We emphasize that the expression in eq.~(\ref{xSec:UNIR_pert}) is given by a product of elementary functions, and thus, numerically much cheaper to calculate.  We stress that {\bf the function in eq.~(\ref{xSec:UNITMD_pert},~\ref{xSec:UNIR_pert}) depends only on the factorization scales  $(\mu_f,\zeta_f)$.}

One of the essential benefits of the optimal TMD definition is that it cuts out the question of the low-energy point normalization. In the suggested universal definition {\bf the low-energy normalization is defined "non-perturbatively" and uniquely by eq.~(\ref{xSec:musaddle}). For that reason  the constant $\mu_i$ is absent together to the associated uncertainty factor $c_3$.} The part of the ambiguity related to the non-ideal perturbation theory is pumped into $c_2$ (since effectively, in eq.~(\ref{xSec:UNIR}), $\mu_f=\mu_i$). Therefore, the error-band for this cross-section can be obtained by the variation of $c_2$ and $c_4$ only. The same is true for the cross-section in the form eq.~(\ref{xSec:UNITMD_pert}). 

The TMD distribution $\tilde F$ is not entirely the optimal TMD distribution. In particular, the coefficient function for small-$b$ matching of $\tilde F$ is given by $C^{\text{pert}}$ defined in eq.~(\ref{th:C_in_zeta}) with $c_{1}=0$ and $c_{2}=\gamma_1 d^{(2,0)}/\Gamma_0$. We note that this definition of $\tilde F$ coincides with the definition of $\tilde F$ in the "naive" $\zeta$-prescription used in ~\cite{Scimemi:2017etj}. 
The formula (\ref{xSec:UNIR_pert}) is very simple for practical implementation, since it has no integration  and does require a solution of eq.~(\ref{xSec:musaddle}), but it consists only of   sums and products of elementary functions. 

At the physical point $(\mu_f,\zeta_f)=(Q,Q^2)$, the expression for TMD distribution reads
\begin{eqnarray}\label{F(Q)}
F_{f \ot h}(x,b;Q,Q^2)=\(C_0 Q b\)^{-\mathcal{D}^f_{\text{NP}}(Q,b)}e^{-\mathcal{D}^f_{\text{NP}}(Q,b)v(Q)}\tilde F_{f\ot h}(x,b).
\end{eqnarray}
In this expression the coupling constant is defined at  \textit{fixed} hard scale $Q$, and it is in principle  small. At very small-$b$ ($b\ll Q^{-1}$) and large-$b$ ($b\gg Q^{-1}$) the  contribution of the logarithms can potentially appear. This behavior is unavoidable, because any resummation procedure that would move the scale inside the logarithm to a better value is equivalent to a redefinition of a point on the null-evolution line, and thus it reduces to an un-evolved expression.

At the leading order, the expression (\ref{F(Q)}) has a very simple explicit form. Indeed, substituting eq.~(\ref{app:DasX},~\ref{app:defZ}) with the leading coefficients defined in eq.~(\ref{app:d0},~\ref{app:g0}) into eq.~(\ref{F(Q)}), we obtain 
\begin{eqnarray}
F_{f \ot h}(x,b;Q,Q^2)=\(\frac{Q b}{C_0}\)^{-\frac{\Gamma_0}{2\beta_0}}\Big[1-2\beta_0a_s(Q)\ln\(\frac{Q b}{C_0}\)\Big]^{\frac{\Gamma_0}{4\beta_0^2a_s(Q)}}\tilde F_{f\ot h}(x,b).
\end{eqnarray}

All three distributions, namely, the general TMD distribution $F(x,b;\mu,\zeta)$ used in eq.~(\ref{xSec:gammaImproved}); the optimal TMD distribution $F(x,b)$ used in eq.~(\ref{xSec:UNITMD}); and the universal TMD distribution defined on perturbative curve $\tilde F(x,b)$ used in eq.~(\ref{xSec:UNITMD_pert}) are related to each other in unique way:
\begin{eqnarray}
F_{f\ot h}(x,b;\mu,\zeta)&=&\exp\Big[-\mathcal{D}_{\text{NP}}^f(\mu,b)\ln\(\frac{\zeta}{\zeta_\mu(b)}\)\Big]F_{f\ot h}(x,b),
\\
F_{f\ot h}(x,b;\mu,\zeta)&=&\exp\Big[-\mathcal{D}_{\text{NP}}^f(\mu,b)\ln\(\frac{\zeta}{\zeta^{\text{pert}}_\mu(b)}\)\Big]\tilde F_{f\ot h}(x,b).
\end{eqnarray}
In these relations the convergence  improves increasing  the  order of the perturbative series, but they are not affected by solution path-dependence effects. Therefore, given the model for $\mathcal{D}_{\text{NP}}$ the comparison of TMD distribution is straightforward.


In figs.~\ref{fig:ATLAS} and \ref{fig:E288} we compare the variation bands obtained from different versions of the cross-section at NNLO. Note, that to compare the bands we use the same model parameters for all plots, which however does not coincide with the best fit values. One can see that the size of the variation band is slightly decreased in comparison to the standard case given in eq.~(\ref{xSec:standard}). This is the effect of the restoration of solution path-independence. The error bands of eq.~(\ref{xSec:gammaImproved}) and eq.~(\ref{xSec:UNITMD_pert}) do not contain the error coming from  the change of the evolution path. In contrast, the error bands of eq.~(\ref{xSec:gammaImproved}) and eq.~(\ref{xSec:UNITMD_pert}) are practically the same since the only difference between these solutions is the point at which the null-evolution line is used. We appreciate that the solution in eq.~(\ref{xSec:UNITMD_pert}) is numerically more stable, since all parameters are well inside the finite region (the numerical artifacts of error bands in the first and the second lines come from the numerical uncertainty of the extremely small $a_s(b^{-1})$ at asymptotically small $b$. The values of $a_s$ are taken from the MMHT package \cite{Harland-Lang:2014zoa}.) A test of the relative convergence of variation bands and central values is not so simple and will be made in future studies. The source of difficulty is the non-perturbative structure of TMD distributions, that plays an important numerical role, and thus should be fit at each pertrubative order separately.

\section{Conclusion}
The existence of a double-scale evolution of non-perturbative hadronic matrix elements poses new questions regarding an efficient implementation of these observables. In this work we have studied in detail the main consequences of double-scale evolution. We have concentrated on the evolution of TMD distributions. Nonetheless,  our methods and conclusions can possibly be adapted to other non-perturbative functions/distributions. The possible areas of extension include jet-observables \cite{Stewart:2013faa}, resummation in momentum space~\cite{Ebert:2016gcn}, double-parton distributions \cite{Diehl:2011yj}.

A consistent and  efficient composition of TMD factorization formalism is fundamental for the precise extraction of non-perturbative function. It is especially important nowadays when big efforts are running to join a large  amount of data coming from  semi-inclusive DIS and  Drell-Yan~\cite{Bacchetta:2017gcc} or when one wants to include new LHC data, characterized  by a great precision \cite{Scimemi:2017etj}. It is time to tackle the problem of  the stability of the matching  of  the perturbative and non-perturbative parts of TMD factorization formalism. In this respect, it is essential to keep in mind the double nature of the  non-perturbative structure of the TMDs. From one side we have non-perturbative corrections in the evolution factor and from the other side we want to explore the intrinsic non-perturbative  content of the TMDs beyond its collinear limit.  The disentanglement of these two non-perturbative effects results to be fundamental in the TMD program.
 
In this work we have discussed the main problem of the double-scale evolution. Namely, the absence of a unique solution within the (unavoidably) truncated perturbation theory. In this way the final implementation of the TMD evolution does depend on the particular choice of integration path in the $(\mu,\zeta)$ plane. We have demonstrated and described that this problem is poorly cured by an increase of the perturbative order. In standard error estimations, this theoretical error is accounted changing  the parameters $\mu_0$ and $\mu_i$ in eq.~(\ref{xSec:c1234}). Within such schemes, a proper definition of these parameters becomes essential for the extraction of the non-perturbative parts, whereas the theory predicts total independence on the scale fixation. The additional horrifying effect of solution ambiguity is the violation of transitivity of TMD evolution. It makes practically very difficult an accurate comparison of fits made in different schemes. The choice of a conventional set of scales  which facilitates the comparison of different fits does not present much practical advantages. We propose here  instead a way to bypass this problem by a forceful restoration of fundamental properties of the evolution at each order of perturbation theory. As a result, the problem of solution path-dependence is not present. Practically it results into a better control of  variation error-band due to the absence of path-dependent uncertainties.

The recognition of double-scale evolution naturally proceed to the idea of $\zeta$-prescription, which consists in the identification of the TMD distributions by the value of evolution potential, rather then by scales $(\mu,\zeta)$. Such identification leads to many natural advantages. The main two of them is the complete elimination of double logarithms from OPE, and the {\bf disentanglement of TMD evolution from modeling of TMD distribution}.  This feature will result essential in the phenomenological study of the non-perturbative content of TMDs. 

We denote as optimal a particular realization of $\zeta$-prescription which is primely characterized by a unique (non-perturbative) definition. An additional benefit is the outstanding simplicity of numerical implementation of TMD factorization within the optimal definition.

The implementation of TMD factorization within $\zeta$-prescription has only two matching scales, $\mu_f$ and $\mu_{\text{OPE}}$. These scales have different physical meaning, and they are the only necessary scales. Notice, in fact, that in more classical approaches all other scales (such as $\mu_i$ and $\mu_0$) are only intermediate scales which do not depend on kinematics or hadronization and serve the purpose to smooth the transition between different regimes. The scales $(\mu_f,\zeta_f)\sim {\cal O}(Q,Q^2)$ are limited by kinematics and characterize the hard subprocess. The scale $\mu_{\rm OPE}\sim q_T$ appears in the re-factorization of TMDs onto collinear distributions, and characterizes the intrinsic distribution scale. 

The optimal solution is implemented in the code \texttt{arTeMiDe}~\cite{web}. Using the \texttt{arTeMiDe} we have performed the test of various implementation of evolution discussed in the paper. The comparison of the approaches is given in fig.~\ref{fig:ATLAS}, \ref{fig:E288}. We indeed observe all theoretically expected result, such as control of error-band in solution independent schemes and similarity of error-band in the $\zeta$-prescription and non-$\zeta$-prescription schemes. We admit the improved timing and numerical stability of the optimal realization of TMD factorization. Altogether it opens the road for the global fit of data well-separated in energy scales, such as Drell-Yan and SIDIS. More phenomenological studies are expected in the future.

\subsection*{Acknowledgments}  
I. S. acknowledges comments from I. Stewart. A.V. acknowledges A.Bacchetta, A.Manashov and M.Radici for stimulating discussions and useful comments.
I.S. is supported by the Spanish MECD grant FPA2016-75654-C2-2-P and the group UPARCOS. 

\appendix

\section{Resummed expressions}
\label{app:Res}

In ref.~\cite{Echevarria:2012pw} the explicitly resummed expression for the rapidity anomalous dimension has been derived. In this appendix we re-derive it using simpler method and also present the resummed expression for $\zeta$-line. These formulas are to be used for the practical implementation of solutions discussed in the article.

The rapidity anomalous dimension $\mathcal{D}$ is the function of $\mu$ and $b$. In the perturbative expansion the parameter $b$ always come in the combination with $\mu$, namely, via logarithms 
\begin{eqnarray}
\mathbf{L}_\mu=\ln\(\frac{\mu^2b^2}{4e^{-2\gamma_E}}\).
\end{eqnarray}
At order $a_s^N$ the perturbative expansion is a polynomial of order $N$ in $\mathbf{L}_\mu$. Order-by-order in perturbation theory it satisfies the equation (\ref{th:dDD=G}), and thus, the elder powers of $\mathbf{L}_\mu$ could be derived from the previous orders.

In ref.~\cite{Echevarria:2012pw} it has been shown that the resummation of logarithms leads to the expression which depends on the parameter
\begin{eqnarray}\label{app:X}
X=\beta_0 a_s(\mu)\mathbf{L}_\mu.
\end{eqnarray}
To derive this function we introduce the partially resummed series
\begin{eqnarray}\label{app:DasX}
\mathcal{D}(\mu,b)=\sum_{n=0}^\infty a_s^n(\mu)d_n(X),
\end{eqnarray}
where $d_n$ is a function of $X$. Substituting this expansion into (\ref{th:dDD=G}) and collecting equal power of $a_s$ we obtain the infinite set of equations for functions $d_n$. They are
\begin{eqnarray}
\beta_0 d_n'-\sum_{k=0}^n \beta_k ((n-k)d_{n-k}+Xd_{n-k}')=\frac{\Gamma_n}{2},
\end{eqnarray}
where we omit the argument $X$ of the functions $d$ for brevity. These equations could be solved recursively starting from the equation at $n=0$ which has the form
\begin{eqnarray}
\beta_0(1-X)d_0'=\frac{\Gamma_0}{2}.
\end{eqnarray}
The boundary condition for equations are
\begin{eqnarray}
d_n(X=0)=d^{(n,0)},
\end{eqnarray}
where $d^{(n,0)}$ are the coefficients of the perturbative expansion defined in (\ref{th:dnk}).

The solutions of these equation are
\begin{eqnarray}\label{app:d0}
d_0(X)&=&-\frac{\Gamma_0}{2\beta_0}\ln(1-X),
\\
d_1(X)&=&\frac{1}{2\beta_0(1-X)}\Big[-\frac{\beta_1\Gamma_0}{\beta_0}(\ln(1-X)+X)+\Gamma_1X\Big],
\\
d_2(X)&=&\frac{1}{(1-X)^2}\Big[\frac{\Gamma_0\beta_1^2}{4\beta_0^3}\(\ln^2(1-X)-X^2\)
+\frac{\beta_1\Gamma_1}{4\beta_0^2}\(X^2-2X-2\ln(1-X)\)+\frac{\Gamma_0\beta_2}{4\beta_0^2}X^2
\\\nn &&-\frac{\Gamma_2}{4\beta_0}X(X-2)+d^{(2,0)}\Big].
\end{eqnarray}
These expressions coincides with ones derives in ~\cite{Echevarria:2012pw}.

The perturbative expressions for the equipotential line (\ref{th:zetaline_PT},~\ref{def:v}) also could be resummed by the same method. The curve is parametrized as
\begin{eqnarray}\label{app:defZ}
\zeta_\mu=\mu^2 e^{-g(\mu,b)},
\end{eqnarray}
where $g(\mu)$ satisfies the equation 
\begin{eqnarray}\label{app:eqnZeta}
\Gamma(\mu)g(\mu,b)-\gamma_V(\mu)=2\mathcal{D}(\mu,b)\(1-\mu^2 \frac{d}{d\mu^2}g(\mu,b)\),
\end{eqnarray}
which follows from (\ref{th:equipotential_eqn_simple}). The first terms of the perturbative solution are given in (\ref{def:v}). To find the resummed expression we denote
\begin{eqnarray}
g(\mu,b)=\frac{1}{a_s(\mu)}\sum_{n=0}^\infty a^n_s(\mu) g_n(X),
\end{eqnarray}
where $X$ is defined in (\ref{app:X}). The substituting this expression into (\ref{app:eqnZeta}) together with (\ref{app:DasX}) and collecting the common powers of $a_s$ we obtain the set of differential equations for $g_n$. The first equation reads
\begin{eqnarray}
2(1-X)\beta_0d_0g_0'+(\Gamma_0+2\beta_0d_0)g_0=2d_0.
\end{eqnarray}
The boundary condition is $g_0(X=0)\sim X/2$. The expression for next equations are more cumbersome, and we do not present them here. The solutions for $g_n$ can be easily obtained. They are
\begin{eqnarray}\label{app:g0}
g_0(X)&=&\frac{1}{\beta_0}\frac{X+\ln(1-X)}{\ln(1-X)},
\\
g_1(X)&=&\frac{\beta_1}{2\beta_0^2}\ln(1-X)-\frac{\beta_1}{\beta_0^2}\frac{X}{(1-X)\ln(1-X)}
+\frac{\beta_0\Gamma_1-\beta_1\Gamma_0}{\beta_0^2\Gamma_0}\frac{X^2}{(1-X)\ln^2(1-X)}+\frac{\beta_0\gamma_1-\Gamma_1}{\beta_0\Gamma_0},
\\
g_2(X)&=&
\frac{\beta_1^2}{2\beta_0^3}\frac{\ln(1-X)}{1-X}+\frac{X}{(1-X)^2\ln(1-X)}\Big[\frac{\beta_1^2}{\beta_0^3}+\frac{\beta_2}{\beta_0^2}(1-X)-\frac{\beta_1\Gamma_1}{\beta_0^2\Gamma_0}(2-X)\Big]
\\&&\nn+\frac{X}{(1-X)\ln(1-X)}\frac{\beta_0\gamma_1\Gamma_1-\Gamma_1^2-\beta_0\gamma_2\Gamma_0+\Gamma_0\Gamma_2}{\beta_0\Gamma_0^2}
+\frac{2d^{(2,0)}}{\Gamma_0(1-X)\ln(1-X)}
\\&&\nn+\frac{X^2}{(1-X)^2\ln^2(1-X)}\frac{\beta_1^2\Gamma_0(4-X)+(X-6)\beta_0\beta_1\Gamma_1-\beta_0^2\Gamma_2 X+\beta_0\beta_2\Gamma_0 X+2\beta_0^2\Gamma_2}{2\beta_0^3\Gamma_0}
\\&&\nn+\frac{X}{(1-X)^2\ln^2(1-X)}\frac{2d^{(2,0)}}{\Gamma_0}+\frac{X^3}{(1-X)^2\ln^3(1-X)}\frac{(\beta_0\Gamma_1-\beta_1\Gamma_0)^2}{\beta_0^3\Gamma_0^2}
\\\nn &&-\frac{\beta_1^2\Gamma_0-2\beta_0\beta_2\Gamma_0+\beta_0\beta_1\Gamma_1}{2\beta_0^3\Gamma_0}
-\frac{\beta_1\Gamma_1}{2(1-X)\beta_0^2\Gamma_0}+\frac{\beta_1^2}{2(1-X)^2\beta_0^3}.
\end{eqnarray}

\bibliography{Refs}

\end{document}